\documentclass[fleqn,usenatbib,onecolumn]{mnras}

\usepackage{newtxtext,newtxmath}

\usepackage[T1]{fontenc}
\usepackage{soul}
\usepackage{mathtools}

\DeclareRobustCommand{\VAN}[3]{#2}
\let\VANthebibliography\thebibliography
\def\thebibliography{\DeclareRobustCommand{\VAN}[3]{##3}\VANthebibliography}

\usepackage{graphicx}	
\usepackage{amsmath}	

\title[]{The modulation effect of ice thickness variations on convection in icy ocean worlds}

\author[Kang et al.]{
  Wanying Kang,$^{1}$\thanks{E-mail: wanying@mit.edu}
\\
$^{1}$Earth, Atmospheric and Planetary Science Department, Massachusetts Institute of Technology, Cambridge, MA 02139, USA
}

\pubyear{2023}

\begin{document}
\label{firstpage}
\pagerange{\pageref{firstpage}--\pageref{lastpage}}
\maketitle

\begin{abstract}
   It has been long puzzling whether the ice thickness variations observed on Enceladus can be sustained sorely by a polar-amplified bottom heating. The key to this question is to understand how the upward heat transport by convective plumes would be interfered by the temperature and salinity variations beneath the ice due to the ice thickness variations, which however, has yet to be explored. Here, we find that the horizontal temperature variation induced by the ice topography can easily be orders of magnitude greater than the vertical temperature variation induced by bottom heating using scaling analysis. Due to the dominance of horizontal temperature gradient, convective plumes are completely shut off by a stratified layer under the thin ice formed out of baroclinic adjustment, largely slowing down the vertical tracer transport. The stratified layer will also deflect almost all of the core-generated heating toward the regions with thicker ice shell, destroying the ice thickness gradient. These results allow us to put an upper bound on the core-generated heating on Enceladus, which is crucial for the estimate of habitability. Scaling laws for the bottom heat flux to penetrate the stratification is derived and examined. This scaling can be used to constrain the maximum ice thickness variations induced by heterogeneous bottom heating on icy satellites in general, which can be used to differentiate icy satellites that generate the majority of heat in the ice shell from those that generate the majority of heat in the silicate core.
\end{abstract}



The partition of heat production between the silicate core and ice shell plays a vital role in determining Enceladus' habitability, because it reflects the activity of hydrothermal reactions and affect the tracer transport timescale \citep{Hsu-Postberg-Sekine-et-al-2015:ongoing, Choblet-Tobie-Sotin-et-al-2017:powering, McKay-Davila-Glein-et-al-2018:enceladus, Kang-Marshall-Mittal-et-al-2022:ocean}. The poleward-thinning ice geometry on Enceladus is generally in line with the polar-amplified tidal dissipation in the ice \citep{Hemingway-Mittal-2019:enceladuss, Kang-Flierl-2020:spontaneous, Kang-Mittal-Bire-et-al-2022:how}, but tidal models for the ice shell are unable to reproduce enough heat to balance the conductive heat loss \citep{Beuthe-2019:enceladuss, Soucek-Behounkova-Cadek-et-al-2019:tidal}, motivating previous studies to look into the possibility for the silicate core to produce heat under tidal deformation \citep{Choblet-Tobie-Sotin-et-al-2017:powering, Liao-Nimmo-Neufeld-2020:heat, Rovira-Navarro-Katz-Liao-et-al-2022:tides}. Due to the remarkable uncertainty associated with the rheology of the ice \citep{Robuchon-Choblet-Tobie-et-al-2010:coupling, Shoji-Hussmann-Kurita-et-al-2013:ice, Behounkova-Tobie-Choblet-et-al-2013:impact, McCarthy-Cooper-2016:tidal, Beuthe-2019:enceladuss, Soucek-Behounkova-Cadek-et-al-2019:tidal, Gevorgyan-Boue-Ragazzo-et-al-2020:andrade}, and even more so, the uncertainty associated with the silicate core rheology \citep{Travis-Schubert-2015:keeping, Choblet-Tobie-Sotin-et-al-2017:powering}, the partition of heat production between the ice shell and the silicate core remains poorly constrained.

The dissipation pattern in the core and the shell are both likely to be polar amplified \citep{Beuthe-2018:enceladuss, Beuthe-2019:enceladuss, Choblet-Tobie-Sotin-et-al-2017:powering, Liao-Nimmo-Neufeld-2020:heat, Rovira-Navarro-Katz-Liao-et-al-2022:tides}, which seems in line with the poleward thinning ice geometry observed on Enceladus. However, there is a fundamental difference between the two heat sources in terms of how they may affect the ice geometry. If heat is mainly produced in the ice, all of the heat production can be used to balance heat loss or to melt ice. To the contrary, if heat is mainly produced in the core, the heat needs to be delivered to the ice by ocean dynamics, during which the heating pattern may be significantly redistributed. It has been shown that, the planetary rotation will concentrate heat either equatorward or poleward depending on the Rayleigh number and Ekman number of the convective system \citep{Soderlund-Schmidt-Wicht-et-al-2014:ocean, Amit-Choblet-Tobie-et-al-2020:cooling, Soderlund-2019:ocean, Bire-Kang-Ramadhan-et-al-2022:exploring, Gastine-Aurnou-2023:latitudinal}.

While these previous works provide useful insights for meridional heat redistribution by ocean capped by a flat isothermal upper boundary, the effect of ice topography on heat redisctribution has been ignored. 
Because Enceladus' ice thickness varies by $\sim 20$-$25$~km from the equatorial to the polar regions \citep{Hemingway-Mittal-2019:enceladuss, McKinnon-Schenk-2021:new}, the temperature at the water-ice interface will vary by $\sim 0.1$-$0.2$~K following the Clausius-Clayperon relationship. The density gradients under the ice can drive overturning circulation and baroclinic eddies in the ocean \citep{Kang-Mittal-Bire-et-al-2022:how, Kang-Bire-Marshall-2022:role, Kang-Jansen-2022:icy, Kang-2022:different}. Both of the overturning cell and eddies transport heat equatorward (downgradient), and in the meanwhile, enhance the ocean's stratification by making dense water slide underneath buoyant water \citep{Vallis-2006:atmospheric, Callies-Ferrari-2018:baroclinic}. As to be shown later, this stratification may impede the convective plumes from directly reaching the ice, substantially modifying the heating pattern as heat is transported upward. Our goal is to understand the interaction between convection and the meridional density variations under a thickness-varying ice shell, and find threshold for bottom heat flux to penetrate the stratified layer.

\section{Ice forcing, bottom heating and the interplay between the two.}
\label{sec:forcings-interplay}
  
\begin{figure*}
    \centering \includegraphics[page=1,width=0.9\textwidth]{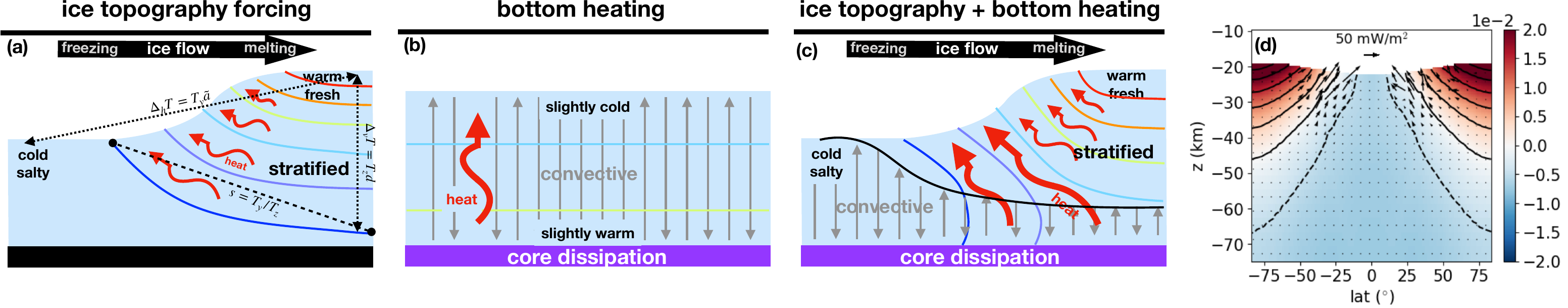}
    \caption{\small{Panel (a-c) sketches the ocean circulation and heat transport driven by ice topography, by bottom heating and by both, respectively. Buoyancy contours are sketched in solid curves, and buoyancy increases from cold colors to warm colors. The buoyancy gradient induced by bottom heating (panel b) is likely much weaker than that forced by the ice topography (panel a) on Enceladus. Grey arrows and curly red arrows represent ocean circulation and ocean heat transport respectively. See main text for details. Panel (d) shows the model diagnosed heat flux in arrows, overlaid on top of temperature in contours and shading, to support the schematics shown in panel a and c.   }}
    \label{fig:schematics}
  \end{figure*}

\textbf{Ocean circulation and heat transport driven by forcings from the ice.} In absence of bottom heating, ocean circulation is sorely driven by ice topography (see Fig.~\ref{fig:schematics}a). On one hand, ice thickness variations would induce pressure gradients at water-ice interface assuming isostasy, and pressure gradients will shift the freezing point following Clausius-Clayperon relationship, leading to temperature variations at the water-ice interface. On the other hand, in order to sustain the ice thickness variations against ice flow, freezing needs to occur in the thick ice regions and vice versa, leading to a salinity/freshwater flux into the ocean \citep{Lobo-Thompson-Vance-et-al-2021:pole, Kang-Mittal-Bire-et-al-2022:how, Ashkenazy-Sayag-Tziperman-2018:dynamics}. As a result, water beneath the thicker part of the ice shell should be colder and saltier than water under a thin ice shell. Both low temperature and high salinity increases water density under the thick ice \footnote{Temperature drop may decrease water density when the water pressure and salinity are sufficiently low. this may be relevant to Enceladus, see \citet{Zeng-Jansen-2021:ocean, Kang-Mittal-Bire-et-al-2022:how, Kang-Bire-Marshall-2022:role} but is not considered here}, creating meridional density gradient in the upper ocean. Although density varies with both temperature and salinity, in this work, we only consider the temperature effect. This simplification would help us focus on how heat forcing from the ice interferes with the heat forcing from the seafloor.

 To provide an order of magnitude estimate, when the ice thickness changes by $\Delta H$, the shift of sub-ice freezing point will be 
\begin{equation}
  \Delta_h T=b_0\Delta P=b_0\rho_i g \Delta H,
  \label{eq:deltahT}
\end{equation}
where $g$ is gravity, $\rho_i$ is ice density and $b_0=7.76\times 10^{-8}$K/Pa is the freezing point suppression coefficient.
Substituting $\Delta H\sim 25$~km and Enceladus gravity, we get $\Delta_hT\sim200$~mK \citep{Hemingway-Mittal-2019:enceladuss, McKinnon-Schenk-2021:new}. On a larger icy moon like Europa, the same $\Delta_hT$ can be achieved by an ice thickness contrast of merely 2~km, due to the stronger gravity. 

The dense water formed under the thick equatorial ice will sink and fill the deep ocean, and the buoyant water formed under the polar thin ice will be diffused downward by the vertical mixing induced by tidal-libration motion \citep{Rekier-Trinh-Triana-et-al-2019:internal, Rovira-Navarro-Matsuyama-Hay-2023:thin}. These two processes sustain a meridional density gradient in the ocean interior (see solid contours in Fig.~\ref{fig:schematics}a), which then drives baroclinic instability \citep{Charney-1947:dynamics, Stone-Hess-Hadlock-et-al-1969:preliminary}.

Properties of baroclinic eddies and the resultant heat transport have been intensely studied in the context of earth ocean and atmosphere \citep[e.g., ][]{Charney-Stern-1962:stability, Held-Larichev-1996:scaling, Jansen-Ferrari-2013:equilibration}. These eddies are known to transport heat upward along the isopycnals (constant density contours) \footnote{This property arises because motions across isopycnals are strongly inhibited, and without diapycnal motion, heat can only be transported along-isopycnal \citep{Jayne-Marotzke-2002:oceanic}.} as sketched in Fig.~\ref{fig:schematics}a and shown in Fig.~\ref{fig:schematics}d. The horizontal component of heat transport $\mathcal{F}_h$ reduces the meridional temperature gradient. The upward heat transport $\mathcal{F}_v$, as manifested by warm water rising and cold water sinking, in turn converts the fluid's gravitational potential energy into kinetic energy at the expense of increasing ocean stratification. In fact, without vertical diffusion continuously steepening the isopycnals, baroclinic would exhaust the potential energy and set the isopycnals completely horizontal \cite{Young-2010:dynamic, Jansen-Kang-Kite-2022:energetics}, after that, the baroclinic instability would stop. With vertical diffusion, a finite isopycnal slope can be maintained, which determines $\mathcal{F}_h$ and $\mathcal{F}_v$. Adapting geostrophic turbulence theory to icy moon oceans, \citep{Kang-2022:different} suggested scalings law for $\mathcal{F}_h$ and $\mathcal{F}_v$, the derivation of which is reviewed in the appendix A.


\textbf{Ocean circulation and heat transport driven by bottom heating.} With sorely bottom heating, the ocean will convect, assuming ocean salinity and pressure are sufficiently high to suppress water's anomalous expansion (i.e., $\alpha>0$). While heat is transported upward, it will be only slightly concentrated equatorward or poleward ($<20\%$) depending on relative strength of rotation, heat flux and viscosity/diffusivity \citep{Amit-Choblet-Tobie-et-al-2020:cooling, Bire-Kang-Ramadhan-et-al-2022:exploring, Soderlund-Schmidt-Wicht-et-al-2014:ocean}. Although the orientation of convective plumes and heat transport are modulated by the planetary rotation \citep{Soderlund-Schmidt-Wicht-et-al-2014:ocean, Gastine-Wicht-Aubert-2016:scaling, Soderlund-2019:ocean, Amit-Choblet-Tobie-et-al-2020:cooling, Bire-Kang-Ramadhan-et-al-2022:exploring}, to the first order, the ocean dynamics is dominated by convective plumes shooting upward as sketched in Fig.~\ref{fig:schematics}b. In equilibrium state, a small but finite vertical temperature gradient $\Delta_v T$ will be sustained. In the fast-rotating non-diffusive limit \citep{Gastine-Wicht-Aubert-2016:scaling}, this $\Delta_v T$ follows
\begin{equation}
  \label{eq:deltavT-Nu}
  \Delta_v T\sim \left(\frac{Q\Omega^2}{0.15\rho C_p}\right)^{2/5}D^{1/5}(\alpha g)^{-3/5}.
\end{equation}
Assuming thermal expansivity $\alpha=10^{-4}$/K, and bottom heat flux $Q=40~$mW/m$^2$, which approximately balances the heat loss through a 20~km ice shell, we get $\Delta_vT\sim 7$~mK for Enceladus ($D=40$~km) and $\Delta_vT\sim 0.4$~mK for Europa ($D=90$~km).

\begin{figure*}
    \centering \includegraphics[page=3,width=0.6\textwidth]{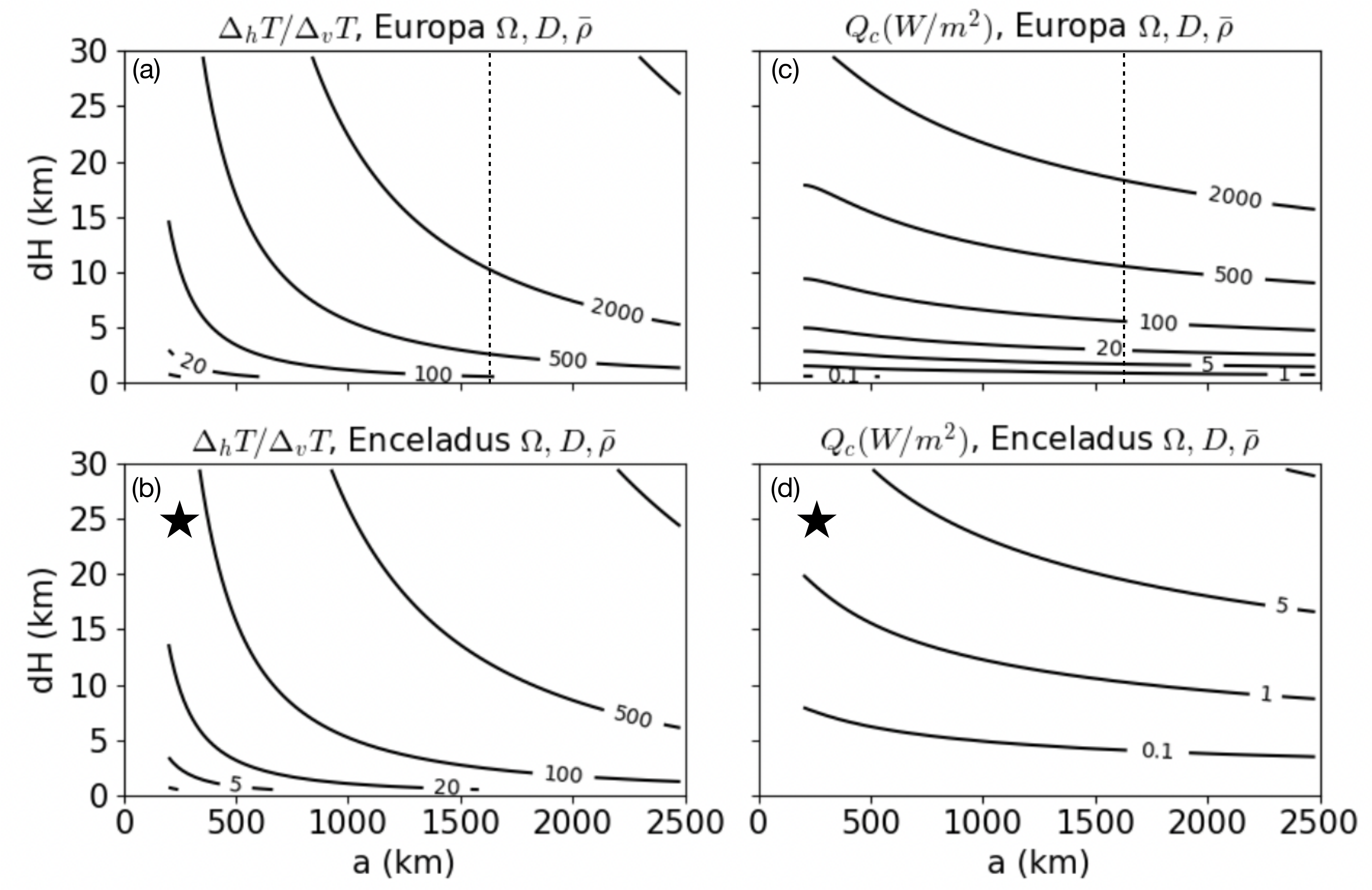}
    \caption{\small{$\Delta_hT/\Delta_vT$ ratio (panel a,b) and critical bottom heat flux $Q_c$ (panel c,d) plotted as a function of satellite radius $a$ and equator-to-pole ice thickness contrast $\Delta H$. Panel (a,c) and (b,d) uses rotation rate, ocean depth and bulk density of Europa and Enceladus, respectively. In panel (a,c), the parameter regime relevant to Enceladus ($a=252$~km, $\Delta H=25$) is marked by a black star, and in panel (b,d), Europa radius is marked by a black dashed line.}}
    \label{fig:Qc}
  \end{figure*}

This vertical temperature gradient $\Delta_vT$ induced by bottom heating, is small compared to the horizontal temperature gradient under the ice $\Delta_hT$ (Eq.~\ref{eq:deltahT}) even with only moderate ice thickness variation. Shown in Fig.~\ref{fig:Qc}(a,b) are the $\Delta_hT/\Delta_vT$ ratio plotted as a function of satellite radius $a$ and ice thickness variation $\Delta H$. In the two panels, Europa and Enceladus' rotation rate $\Omega$, ocean depth $D$ and mean density $\bar{\rho}$ are adopted, respectively. Thermal expansivity $\alpha$ is set to $10^{-4}$/K and bottom heating $Q$ is set to 40~mW/m$^2$ to maximize $\Delta_vT$. As can be seen, the $\Delta_hT$ is 1-3 orders of magnitude greater than $\Delta_vT$ for most of the parameter space explored here. Due to Europa's slow rotation, deep ocean depth and high mean density, the $\Delta_hT/\Delta_vT$ ratio is even higher compared to equivalent scenarios using Enceladus' parameters. With $a=250$~km and $\Delta H=25$~km, Enceladus' $\Delta_hT/\Delta_vT=30$. Europa's ice geometry is not well understood \citep{Nimmo-Thomas-Pappalardo-et-al-2007:global}, leaving $\Delta_hT/\Delta_vT$ ratio poorly constrained. However, because Europa gravity is much stronger, $\Delta_hT$ is smaller than $\Delta_vT$ only when its ice thickness variation $\Delta H<4$ meters. Therefore, unless convection is completely suppressed by anomalous expansion near the surface \citep{Melosh-Ekholm-Showman-et-al-2004:temperature, Vance-Goodman-2009:oceanography, Zeng-Jansen-2021:ocean}, $\Delta_h T$ is likely to be greater than $\Delta_v T$.

\textbf{Ocean circulation and heat transport when both forcings are present.} With both ice topography and bottom heating (Fig.~\ref{fig:schematics}c), the ocean under the thick ice would remain convectively unstable, driven by buoyancy sink at the top and buoyancy source at the bottom. However, under the thin ice shell, the temperature of the water-ice interface there is warmer than that in the thick ice latitudes by $\Delta_hT$. To allow convective plumes to reach the ice shell, the temperature at the seafloor needs to be even higher \footnote{Salinity-induced density anomalies will tend to suppress convection, because to balance the flattening effect of ice flow, the thinner part of the ice shell needs to melt, and the freshwater produced during the melting will stratify the upper part of the ocean \citep{Lobo-Thompson-Vance-et-al-2021:pole, Kang-Mittal-Bire-et-al-2022:how}.}. In other words, the meridional temperature gradient $\Delta_hT$ (\eqref{eq:deltahT}) set by the ice topography needs to extend throughout the entire ocean depth. Having such strong meridional temperature gradient prevail the entire ocean will trigger vigorous baroclinic instability (or slantwise convection) even in presence of convection, redistributing heat down-gradient, and meanwhile, restratifying the upper ocean and suppressing convection \citep{Callies-Ferrari-2018:baroclinic}.

If $\Delta_hT$ is large enough, a stratified layer will form in the upper ocean under the thin ice, preventing convective plumes and its heat from reaching the thinner part of the ice. In Fig.~\ref{fig:schematics}c, we demarcate this stratified layer from the convective layer using a black solid curve. In the convective zone, heat is transported straight upward by convective plumes. However, after reaching the stratified layer, baroclinic eddies become the vehicle of upward heat transport in place of convective plumes. Unlike the convective layer, the heat transport by baroclinic eddies is aligned with the direction of the isopycnals \citep[][,see also appendix A]{Jayne-Marotzke-2002:oceanic}. This property of baroclinic eddies can potentially give rise to significant lateral heat redistribution as heat is transported upward. Whether heat can reach the thinnest part of the ice shell depends on the slope of the isopycnals. Only when all isopycnals originated from the ice are steep enough to touch the seafloor, the heat source, can heat be delivered to the entire ice shell. If any isopycnal originated under the thin ice region does not touch the seafloor, then hardly any heat can reach there. Instead, almost all the bottom heating will be deflected toward the thick ice regions, as sketched in Fig.~\ref{fig:schematics}c. 

The transition from complete to incomplete heat deflection occurs when isopycnals originated from the ice around mid-latitude touches the seafloor. This criterion can be rewritten as isopycnal slope $s=D/\tilde{a}$ ($D$ denotes ocean depth, and $\tilde{a}=a-(H_0+D)/2$ is the radius at the mid-depth of the ocean-ice layer, $H_0$ is the mean ice thickness), ignoring the depth of the convective lower layer. With $s<D/\tilde{a}$ almost all of the bottom heating will be focused toward the equator, where ice is thick, following the direction of isopycnals (see schematics Fig.~\ref{fig:schematics}c). In this case, the horizontal heat transport $\mathcal{F}_h$ equals to the vertical heat transport $\mathcal{F}_v$ as shown in the appendix A, meaning that all of the bottom heat released from the polar seafloor will be deflected equatorward. With $s>D/\tilde{a}$, more and more isopycnals penetrate deep enough to touch the seafloor, and $\mathcal{F}_h$ starts to become smaller than $\mathcal{F}_v$ (see appendix A for derivation), allowing part of the bottom heating to be not deflected. At the transitional point ($s=D/\tilde{a}$), the upward heat transport $\mathcal{F}_v$ by baroclinic eddies can be written as
\begin{equation}
  \label{eq:critical-F}
  \left.\mathcal{F}_v\right|_{s=D/\tilde{a}}=\left. \frac{2 \pi k C_{p} \rho_{0}}{ f^{2}}(\alpha g \tilde{a})^{\frac{3}{2}} (\Delta_h T s)^{\frac{5}{2}}\right|_{s=D/\tilde{a}}=\frac{2 \pi k C_{p} \rho_{0}}{ f^{2}\tilde{a}}(\alpha g )^{\frac{3}{2}} (\Delta_h T D)^{\frac{5}{2}}.
\end{equation}
 In an equilibrium state, the upward heat transport by baroclinic eddies plus the downward heat diffusion in the stratified upper ocean needs to equal to the prescribed heat flux at the seafloor to avoid heat accumulation in the interior. This heat flux balance allows us to estimate the bottom heat flux $Q_c$ needed to sustain an isopycnal slope of $s=D/\tilde{a}$,
\begin{eqnarray}
  Q_c&=&\frac{\left.\mathcal{F}_v\right|_{s=D/\tilde{a}}}{2\pi \tilde{a}^2}-\kappa_vC_p\rho_0\frac{\Delta_v T}{D}\nonumber\\
  &=&\frac{k C_{p} \rho_{0}}{\tilde{a}^3 f^{2}}(\alpha g)^{\frac{3}{2}} (\Delta_h T D)^{\frac{5}{2}}-\kappa_vC_p\rho_0\frac{\Delta_h T}{D} \nonumber\\
     &\approx& \frac{k C_{p} \rho_{0}}{\tilde{a}^3 f^{2}}(\alpha g)^{\frac{3}{2}} (\Delta_h T D)^{\frac{5}{2}}.
                 \label{eq:Qc}
\end{eqnarray}
As bottom heat flux increases beyond $Q_c$, we expect the heat deflection to transition from complete to incomplete. From step 1 to step 2, we substituted $\Delta_hT=\Delta_vT$, which holds when isopycnals just start to touch the seafloor. The contribution by diffusion (2nd term above) is dropped, because it is generally much smaller. To make it comparable with the contribution by bottom heating (1st term above), $\kappa_v$ needs to be comparable to $\kappa_{vc}\equiv k (\alpha g)^{\frac{3}{2}} \Delta_h T^{\frac{3}{2}} D^{\frac{7}{2}}/(\tilde{a}^3 f^{2})$, which in turn can be rewritten into $k (4\pi G\rho_b/3)^3(b_0\alpha\rho_i\Delta H)^{\frac{3}{2}} D^{\frac{7}{2}}/f^2$ using Eq.~\eqref{eq:deltahT} and $g=(4\pi/3)G\rho_b\tilde{a}$, where $G$ is the gravitational constant, $\rho_b$ is the bulk density of the satellite. The main factors that determine this quantity are $D$, $\Delta H$ and $f$. If we substitute, $D=40$~km, $\Delta H=2$~km, $f=10^{-4}$/s and $\rho_b=3000$~kg/m$^3$, we get a $\kappa_{vc}$ of $0.01$~m$^2$/s, which is 1-2 orders of magnitude greater than the maximum $\kappa_v$ so far has been predicted for icy satellites \citep{Rekier-Trinh-Triana-et-al-2019:internal, Rovira-Navarro-Matsuyama-Hay-2023:thin}. This is also consistent with \citet{Kang-2022:different} in that, with $\kappa_v$ set to $10^{-3}$~m$^2$/s, the isopycnals in the simulations only reach the mid-depth of the ocean.

Fig.~\ref{fig:Qc}(c,d) shows how $Q_c$ varies with satellite radius $a$ and equator-to-pole ice thickness contrast $\Delta H$. Rotation rate $\Omega$, ocean depth $D$ and bulk density $\bar{\rho}$ of Europa and Enceladus are assumed in panel (c) and (d), respectively; and $\alpha$ is set to $10^{-4}$/K. As can be seen, $Q_c$ are generally much greater than the $O(10)$~mW/m$^2$ total dissipation rate expected for an icy moon with an $O(10)$~km ice shell, and $Q_c$ increases with the icy moon's size and ice topography. Using Enceladus' $\Omega,\ D,\ \bar{\rho}$, $Q_c$ can easily be several W/m$^2$. Due to the slow rotation, likely deep ocean and high bulk density of Europa, $Q_c$ easily reaches hundreds or even thousands of W/m$^2$ assuming an ice thickness variation of 3~km, and in order to keep $Q_c$ below $40$~mW/m$^2$, the ice thickness variation needs to be smaller than 140~m.

For Enceladus, in particular, if we set $\Delta H$ to 25~km and assume $\alpha=10^{-4}$/K \footnote{We choose a relatively large $\alpha$ to account for the density gradient induced by salinity anomalies \citep{Kang-Mittal-Bire-et-al-2022:how}.}, the critical heat flux $Q_c=2$~W/m$^2$. Such a bottom heat flux is over an order of magnitude greater than the estimated conductive heat loss rate of $40$~mW/m$^2$ based on the observed mean ice thickness of $\sim$20~km \citep{Beuthe-Rivoldini-Trinh-2016:enceladuss, Hemingway-Mittal-2019:enceladuss}. Even if all heat is produced in the silicate core over 30\% of the surface area over the poles, the resultant heat flux per area is still an order of magnitude smaller than $Q_c$.
It has been proposed that the heat flux at the seafloor could be highly unevenly distributed and local heat flux can reach $\sim 10$~W/m$^2$ \citep{Choblet-Tobie-Sotin-et-al-2017:powering}, however, as demonstrated by \citet{Kang-Marshall-Mittal-et-al-2022:ocean}, focused bottom heating will trigger baroclinic instability along the boundary that separates the warm plume water and the cold ambient water, and the resultant turbulence efficiently mixes the heat laterally within a few kilometers above the seafloor. Therefore, by the time heat reaches the surface stratified layer, the local inhomogeneity should have already been removed.

 \section{Model configurations.}
  \label{sec:model-configuration}
  To examine the prediction (section~\ref{sec:forcings-interplay}) about the interference between the bottom heating forcing and ice topography forcing, we conduct a series of 3D numerical simulations using MITgcm \citep{Marshall-Adcroft-Hill-et-al-1997:finite}, where deep shell effect and a full treatment of Coriolis effect are fully accounted for. To simplify the interpretation of the results, we here adopt ``LINEAR'' equation of state and set the saline contractivity $\beta$ to zero. As a result, salinity becomes a passive tracer in our default configuration. At the water-ice interface, temperature is relaxed toward the local freezing point at a fixed rate $\gamma_T$, and the salinity flux is prescribed such that it can counterbalance the tendency induced by the ice flow. The ice thickness profile is set to $H=H_0-H_2 P_2(\sin\phi)$, where the mean ice thickness $H_0=24$~km, $H_2=3$~km, $P_2$ denotes the 2nd order Legendre polynomial and $\phi$ denotes latitude. With $H_0=24$-km, the conductive heat loss rate $\mathcal{H}_{\mathrm{cond}}$ should be around $40$~mW/m$^2$ on global average (Eq.~B10). From equator to the pole, the ice thickness $H$ decreases by $4.5$~km, resulting in an poleward increase of freezing point following Eq.~(\ref{eq:deltahT}). At the bottom of the ocean, a slightly poleward-amplified bottom heating is prescribed. 

To see whether $Q_c$ (Eq.~\ref{eq:Qc}) indeed captures the transition from complete heat deflection to incomplete, we will conduct experiments forced by different levels of bottom heating $Q_0$, at three different satellite radii $a=250$~km, $1000$~km and $2500$~km. As the size of the satellite increases, the critical bottom heat flux $Q_c$ (Eq.~\ref{eq:Qc}) rises from $0.21$~W/m$^2$ to $3.36$~W/m$^2$, and to $17.7$~W/m$^2$ \footnote{In the calculation, thermal expansivity $\alpha$ is set to $8.45\times 10^{-5}$/K, $1.05\times 10^{-4}$/K and $1.46\times 10^{-4}$/K, respectively, corresponding to an assumed salinity of 60~psu and a depth of 24~km.}. This spread would allow us to test whether the proposed criterion is generally applicable.

For each satellite radius, we will investigate five scenarios: 1) the shell-heating scenario ($Q=0$), 2) the core-heating scenario ($Q=\mathcal{H}_{\mathrm{cond}}f_{\mathrm{deep}}$), 3) a subcritical scenario ($Q=0.33Q_c$), 4) the marginally supercritical scenario ($Q=Q_c$) and 5) a highly supercritical scenario ($Q=5Q_c$). When determining $Q$ for scenario 2), we account for the surface area difference at the seafloor and the water-ice interface by including a deep factor $f_{\mathrm{deep}}=((a-H_0)/(a-H_{\mathrm{tot}}))^2$, where $H_0=24$~km is the mean ice thickness and $H_{\mathrm{tot}}=76$~km is the thickness of the ice-ocean layer in total. By conducting the first two experiments, we aim to understand whether bottom heat flux can be transmitted to the thin polar ice shell without violating the global heat balance. Scenarios 3)-5) are used to test the validity of the proposed criterion, as described in \eqref{eq:Qc}, which is intended to distinguish between the regimes with and without a surface stratified layer and substantial equatorward heat convergence. When $a=250$~km, the bottom heating in scenario 2) happens to be very close to that in scenario 3) (difference less than 3\%), so we combined the two scenarios to one.

Each simulation is integrated until thermal equilibrium, i.e., the convergence of meridional ocean heat transport (OHT) plus the bottom heating matches the heat flux deposited to the ice shell. To keep the computational cost manageable, we first spin up each simulation using coarse resolution 2D model, then fine resolution 2D model before starting the 3D simulation. To further accelerate the convergence of the 3D simulations, we pause the integration every week (wall-clock time) and advance the temperature profile using the time-averaged tendency calculated from the simulation during the preceding week. Other model setups are the same as what is used in \cite{Kang-2022:different}. A thorough description of model setup can be found in the Appendix B, and parameters adopted in this study are summarized in Table~B1.

We note that the resolution adopted here is sufficient for baroclinic eddies (characterized by deformation radius) but may not be sufficient for convection. The size of convective plumes is expected to follow $l_{\mathrm{rot}}\sim\sqrt{B/f^3}$, where $B=\alpha g Q_0/(\rho C_p)$ is the bottom buoyancy flux, according to the scalings given by \cite{Jones-Marshall-1993:convection}. E.g., for the three bottom-heated cases with $a=250$~km, $l_{\mathrm{rot}}\sim 0.45,\ 0.77, 1.7$~meters, which is impossible to resolve in global simulations. On the other hand, the cone scaling $l_{\mathrm{cone}}\sim \sqrt{l_{\mathrm{rot}} D}$, which governs the size of the aggregated convective vertices, is 500~m, 700~m and 1000~m for the three bottom-heated cases, and is marginally approachable (our horizontal resolution is 1000~m at the equator and $<$200~m over polar regions). 

At the current resolution, we may underestimate the convective heat transport, and it is thus necessary to test the sensitivity of our results to convective heat transport efficiency. Taking advantage of the scale separation between the convective plumes and the baroclinic eddies ($\sim 10$~km for $a=250$~km cases), we can parameterize the small-scale convection by enhancing local vertical diffusivity, which has been shown to effectively represent the mixing induced by convection \citep{Jones-Marshall-1993:convection}. More complicated parameterizations such as the nonlocal K-Profile Parameterization \citep[KPP][]{Large-McWilliams-Doney-1994:oceanic} has been developed and broadly applied to study earth ocean, but for sake of keeping the convective parameterization transport, we decide not to use those schemes.

  Our default simulations do not have any convective parameterization, and as a result, convection happens at grid scale over the polar regions, especially in the intermediately heated cases ($Q_0=Q_c/3,\ Q_c$). The under-resolved convection may affect its capability to overcome the stratification and thereby heat deflection. To test the sensitivity to convection closure, we repeat the three bottom-heated $a=250$~km experiments with a convective parameterization, which vertically mixes the convectively unstable regions with an enhanced diffusivity of $\kappa_{\mathrm{conv}}=0.38,\ 0.74$ and $1.9$~m$^2$/s, respectively. The $a=250$~km experiments are chosen because their $\Delta_vT/\Delta_hT$ ratio is the largest there, resulting in convective mixing efficiency playing the most significant role. These $\kappa_{\mathrm{conv}}$ are chosen to be $2$ times the following convective mixing rates predicted by scaling laws \citep{Gastine-Wicht-Aubert-2016:scaling},
  \begin{equation}
    \label{eq:kappa-conv}
    \kappa_{\mathrm{conv}}=0.15 (\alpha g \Delta_vT)^{\frac{3}{2}} D^{\frac{1}{2}}\Omega^{-2}=(0.15)^{\frac{2}{5}} \left(\frac{\alpha g Q}{\rho C_p}\right)^{\frac{3}{5}}\left(\frac{D}{\Omega}\right)^{\frac{4}{5}}.
  \end{equation}
  The second equality above is done by substituting \eqref{eq:deltavT-Nu}. The predicted $\kappa_{\mathrm{conv}}$ for the three bottom-forced $250$~km experiments are $0.19$, $0.37$, and $0.96$~m$^2$/s.
  With $\kappa_{\mathrm{conv}}$ equals two times the predicted values, we hope to bracket the true solution in between.

Furthermore, we also test the sensitivity to the water-ice heat exchange coefficient $\gamma_T$, which should be determined by the poorly constrained surface roughness, stratification, shear etc. \citep{Monin-Obukhov-1954:basic}. By default, $\gamma_T$ is set to $10^{-5}$~m/s. Using such a small $\gamma_T$ necessarily means that a relatively large temperature contrast between the water and the ice is needed to transport the certain amount of heat. If the heat is focused equatorward (as shown later), the heat accumulated under the equatorial ice shell will reduce the meridional temperature gradient, limiting the ice topography's capability to deflect heat equatorward. By choosing a small $\gamma_T$, we again stay in the conservative end in estimating the deflected heat flux. To quantify the sensitivity, we repeat all of the strongly forced experiments ($Q_0=Q_c/3,\ Q_c,\ 5Q_c$) with a 50 times larger $\gamma_T$, which may be considered as an upper bound.

  The solutions of these sensitivity tests are presented in the supplementary material section~3.

  \section{Ocean dynamics}
  \label{sec:ocean-dynamics}

  \begin{figure*}
    \centering \includegraphics[page=5,width=\textwidth]{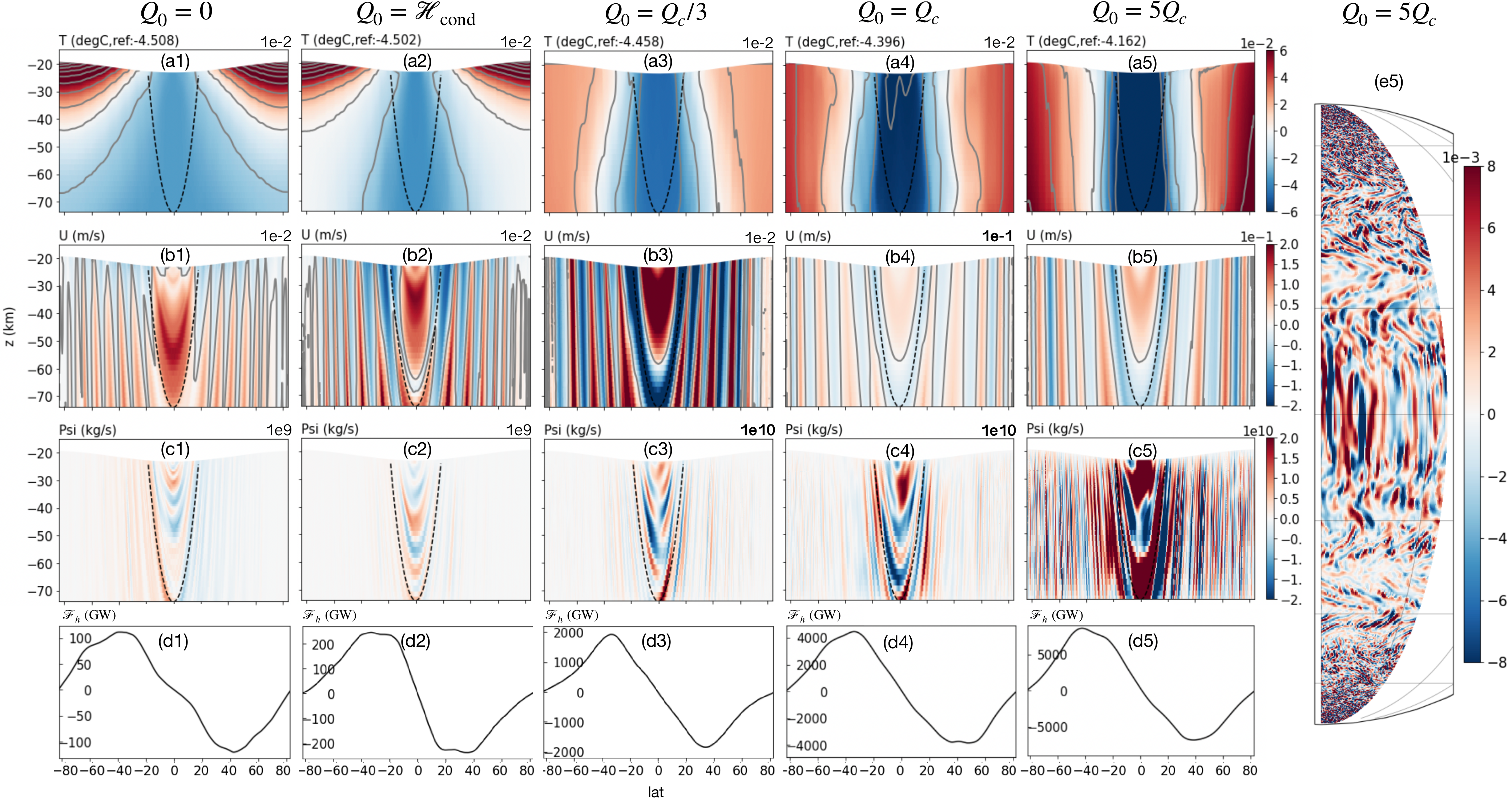}
    \caption{\small{Solutions for the various heating scenarios with $a=1000$~km. The top three rows show zonally-averaged temperature $T$, zonal flow speed $U$, meridional streamfunction $\Psi(\phi,z) = \int_{-H_{\mathrm{tot}}}^z  \rho(\phi,z') V(\phi,z')\times (2\pi(a+z')\cos\phi)~ dz'$, where $\phi$ denotes latitude, $z$ denotes altitude, $-H_{\mathrm{tot}}$ is the altitude of the seafloor and $V$ and $\rho$ are meridional speed and density. $\Psi>0$ indicates clockwise circulation. The row-d shows the vertically-integrated meridional OHT $\mathcal{F}(\phi) = \int_{-H_{\mathrm{tot}}}^{-H}  \rho(\phi,z') V(\phi,z')T(\phi,z')\times (2\pi(a+z')\cos\phi)~ dz'$. $\mathcal{F}>0$ if heat is transport northward. From the left to the right column show solutions with $Q_0=0$, $Q_0=\mathcal{H}_{\mathrm{cond}}f_{\mathrm{deep}}$, $Q_c/3$, $Q_0=Q_c$ and $Q_0=5Q_c$, respectively, where $Q_c=3.36$~W/m$^2$. Panel (e5) shows the mid-level vertical velocity field for the $Q_0=5Q_c$ scenario.}}
    \label{fig:solution-a1000}
  \end{figure*}

  Shown in Fig.~\ref{fig:solution-a1000} are the numerical solutions for the various heating scenarios with $a=1000$~km. In absence of bottom heating (column-1), the entire deep ocean is filled by the cold water formed under the equatorial ice shell, creating a stratified layer over high latitudes (panel-a1). As required by the thermal wind balance, 
  \begin{equation}
    \label{eq:thermal-wind}
    2\mathbf{\Omega}\cdot \nabla U=\frac{1}{a}\frac{\partial b}{\partial \phi},
  \end{equation}
  the zonal flow $U$ changes direction from retrograde to prograde as $z$ increases as can be seen in panel-b1. Here $\mathbf{\Omega}$ is the planetary rotation vector, $b=\alpha g T$ is buoyancy, $\alpha$ is thermal expansivity, $g$ is gravity, $T$ is temperature and $\phi$ denotes latitude. This vertical shear may be able to drive non-synchronized rotation of the ice shell \citep{Ashkenazy-Tziperman-Nimmo-2023:non}. The meridional overturning circulation sinks near the tangent cylinder \footnote{Tangent cylinder is a cylinder whose sides are parallel to the moon's rotation axis and are tangential to the ocean's floor at the equator} and rises near the poles as can be seen in Fig.~\ref{fig:solution-a1000}-c1. This circulation is aligned with the rotation axis in the interior to avoid the zonal acceleration required by angular momentum conservation as a water parcel moves toward or away from the rotation axis; only along the rough boundaries at the top and the bottom, water can flow radially without gaining too much zonal momentum \citep{Kang-Jansen-2022:icy}. The magnitude of this circulation induced by boundary Ekman transport depends on the surface momentum drag coefficient, which is set to $\gamma_M=10^{-4}$~m/s. Momentum transport by eddies helps form the smaller scale circulations and jets, which are evident in Fig.~\ref{fig:solution-a1000}-b,c. The distance between these jets are consistent with Rhines scale (not shown here) as demonstrated by similar experiments in \citet{Kang-2022:different}. Both overturning circulation and baroclinic eddies (not shown here because they are very similar to what is presented in \citet{Kang-2022:different}) transport heat equatorward (down-gradient), and the contribution from eddies dominate by roughly an order of magnitude, consistent with \citet{Kang-2022:different}.

  As the bottom heating strengthens, the stratification induced by baroclinic eddies is gradually eroded. This can be seen by comparing the different panels in row-a of Fig.~\ref{fig:solution-a1000}. When $Q_0=5Q_c$, the entire ocean becomes convectively unstable, allowing heat to be delivered to the polar ice shell (Fig.~\ref{fig:heat-transport}c, black curve). Since the bottom heating enter the energetics as an energy source term \citep{Jansen-Kang-Kite-2022:energetics}, both zonal flow and meridional circulation are significantly strengthened with $Q_0$, and the jets and circulation cells widen following the prediction of Rhines scale (see row-b,c of Fig.~\ref{fig:solution-a1000}). However, the erosion of the polar stratification comes along with the increase of isopycnal slopes (i.e., contours of constant density). As the isopycnals steepen, the meridional ocean heat transport should also increase following the first equality in Eq.~(\ref{eq:critical-F}). This can be clearly seen by comparing different panels in the row-d of Fig.~\ref{fig:solution-a1000}.

Similar plots for the other two satellite radii can be found in the Fig.C1-C2. The temperature and circulation patterns as well as their response to the progressively enhanced bottom heating are qualitatively similar to what is shown in Fig.~\ref{fig:solution-a1000}. However, due to the stronger gravity on larger icy moons, the meridional temperature gradient $\Delta_hT$ is larger there \eqref{eq:deltahT}, and as a result, the zonal flow speed also strengthens following thermal wind balance (Eq.\ref{eq:thermal-wind}). Circulation and ocean heat transport also strengthens as a result of stronger meridional density gradient and the stronger gravity. All these are in consistency with \citet{Kang-2022:different} and \citet{Kang-Jansen-2022:icy}.

\section{Deflection of bottom heating by ice topography. }
\label{sec:heat-deflection}
\begin{figure*}
    \centering \includegraphics[page=4,width=0.8\textwidth]{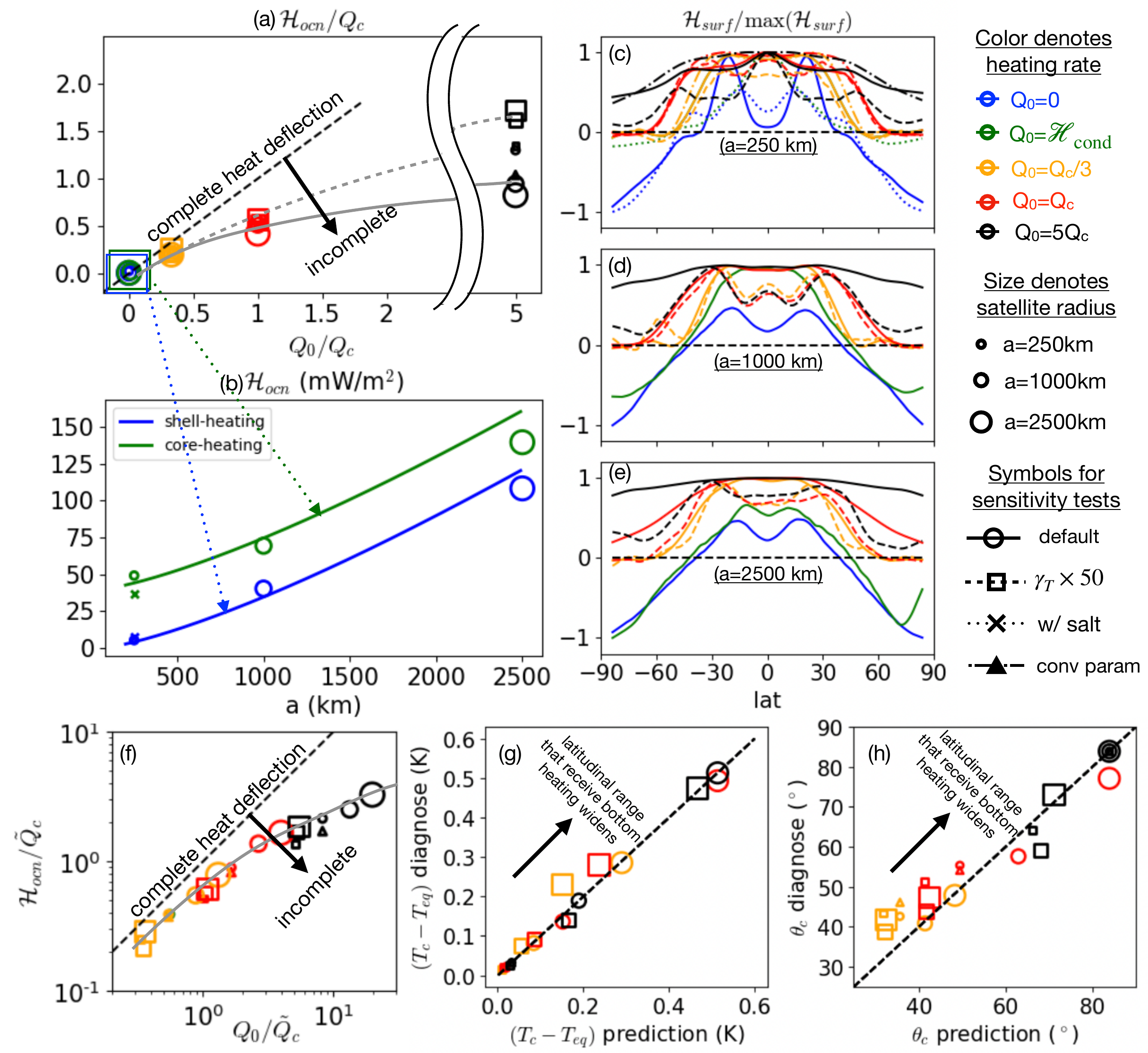}
    \caption{\small{Ocean-ice heat exchange rate induced by meridional OHT $\mathcal{F}_h$. Panel (a) shows $\mathcal{H}_{\mathrm{ocn}}\equiv 2 \overline{|\mathcal{F}_h|}/(\pi a^2)$ normalized by $Q_c$ (Eq.~\ref{eq:Qc}) as a function of the normalized bottom heating $Q_0/Q_c$. The black dashed curve is the one-to-one line. Note the sudden jump from 2 to 5 in the x-axis. Panel (b) shows how $\mathcal{H}_{\mathrm{ocn}}$ varies with the satellite radius $a$ for the shell-heating scenario ($Q=0$) and the core-heating scenario ($Q=\mathcal{H}_{\mathrm{cond}}f_{\mathrm{deep}}$) because they cannot be clearly seen in panel (a) due to their small $Q_0$ relative to the other strongly forced scenarios. The blue curve shows $\mathcal{H}_{\mathrm{ocn}}$ predicted by the scaling (Eq.~\ref{eq:OHT-scaling-shellheating}), and the green curve shows that prediction plus the $Q_0$ prescribed in the core-heating scenario. Panel (c-e) shows the shape of the water-ice heat exchange rate of the five heating scenarios, for $a=250,\ 1000,\ 2500$~km respectively. Panel (f) is similar to panel (a) except that $\mathcal{H}_{\mathrm{ocn}}$ and $Q_0$ are normalized by $\tilde{Q}_c$ solved from Eq.~(\ref{eq:Qc-gamma-modified}), and log-scales are used. Panel (g,h) compare the diagnosed and predicted $T_c-T_{eq}$ and $\theta_c$, where $T_c$ is the sub-ice temperature at the latitude $\theta_c$, which demarcates regions that receive bottom heating from those that do not, and $T_{eq}$ is the sub-ice temperature at the equator.
        In all panels, color is used to differentiate different heating scenarios, and size of the scatter symbols is used to distinguish the three satellite radii. Sensitivity to salt-driven circulation (nonlinear equation of state) is tested for $a=250$~km core/shell-heating cases as shown by cross symbols in panel (b). Sensitivity to parameterized convection is tested for $a=250$~km strongly forced cases as shown by upward triangles in panel (a). Sensitivity to boundary heat exchange coefficient $\gamma_T$ is tested for all strongly forced cases, as shown by square symbols in panel (a). Gray solid and dashed curves are the best fitting curves for the control experiments and the ones with enhanced $\gamma_T$. Panel (c-e) show the shape of the water-ice heat exchange rate of the five heating scenarios (denoted by different colors), for $a=250,\ 1000,\ 2500$~km, respectively. Solid curves show results from default configuration, dashed curves show results with enhanced $\gamma_T$, and dashdot and dotted curves show results with convective parameterization and with salinity effect respectively (only in panel c).  }}
    \label{fig:heat-transport}
  \end{figure*}

Although the absolute magnitude of meridional heat transport $\mathcal{F}_h$ increases with $Q_0$, as can be seen from row-d of Fig.~\ref{fig:solution-a1000}, the percentage of bottom heating that is deflected equatorward should decrease, as the system transitions toward a regime where convection is no longer affected by the meridional temperature gradient under the ice. As suggested by the arguments in section~\ref{sec:forcings-interplay} (see also \citet{Callies-Ferrari-2018:baroclinic}), we expect the transition to occur at $Q_0\sim Q_c$. To see this, we diagnose the globally averaged magnitude of OHT $\overline{|\mathcal{F}_h|}$, multiply it by a factor of $2$ and divide it by the surface area of half-hemisphere $\pi a^2$, to obtain the ocean-ice heat exchange rate $\mathcal{H}_{\mathrm{ocn}}$ induced by this meridional OHT. This $\mathcal{H}_{\mathrm{ocn}}$ is plotted as a function of $Q_0$ in Fig.~\ref{fig:heat-transport}a, after normalizing both by $Q_c$.

Among the five heating scenarios, the three subcritical scenarios ($Q_0=0, \mathcal{H}_{\mathrm{cond}}f_{\mathrm{deep}}, Q_c/3$), all have $\mathcal{H}_{\mathrm{ocn}}$ close to $Q_0$, as evidenced by the proximity of their scatter symbols (blue, green and yellow circles) to the one-to-one line; whilst the other two supercritical scenarios ($Q_0=Q_c,\ 5Q_c$, red and black circles) have $\mathcal{H}_{\mathrm{ocn}}$ significantly smaller than $Q_0$. Despite the fact that $Q_c$ for the largest icy moon considered ($a=2500$~km) is almost 2 orders of magnitude greater than that of the smallest ($a=250$~km), normalizing $Q_0$ and $\mathcal{H}_{\mathrm{ocn}}$ with $Q_c$, reasonably well collapses the three sets of data obtained from experiments with different satellite radii, indicating that $Q_c$ (Eq.~\ref{eq:Qc}) indeed demarcates the regime where bottom heating can reach the polar ice shell from the regime where it cannot. This transition is consistently reflected by the patterns of the upward ocean heat flux shown by solid curves in Fig.~\ref{fig:heat-transport}c-e. Among all experiments, only the experiments with $Q_0=5Q_c$ and the experiment with $a=2500$~km, $Q_0=Q_c$ can deliver heat toward the thin ice over polar regions. Even in those scenarios, the heat flux received by the ice still peaks near the equator, drastically different from the poleward-amplified heating pattern prescribed at the seafloor. 

Since $Q_c$ is 2-4 orders of magnitude greater than the largest possible global-mean bottom heat flux that can be in balance with $\mathcal{H}_{\mathrm{cond}}$, the conductive heat loss rate through the ice shell, the shell-heating and core-heating scenarios are hardly distinguishable from the origin point in Fig.~\ref{fig:heat-transport}a. To better show the ocean heat transport for these cases, we plot $\mathcal{H}_{\mathrm{ocn}}\equiv 2\overline{|\mathcal{F}_h}|/(\pi a^2)$ as a function of the satellite radius $a$ in Fig.~\ref{fig:heat-transport}b in blue and green circles, respectively. The blue curve shows the predicted $\mathcal{H}_{\mathrm{ocn}}$ from the scaling law (Eq.~\ref{eq:OHT-scaling-shellheating}) for the shell-heating scenario, whereas the green curve is the same as the blue curve except shifted upward by $Q_0=\mathcal{H}_{\mathrm{cond}}f_{\mathrm{deep}}$. Such a shift corresponds to having all of the core-generated heating deflected toward the equator. The match between the diagnosed and predicted $\mathcal{H}_{\mathrm{ocn}}$ indicates 1) that the scaling law for the shell-heating scenario given by \citet{Kang-2022:different} works and 2) that hardly any of the bottom heating can reach the polar ice shell in the core-heating scenario.

It has been proposed that a polar-amplified heat flux from the ocean is necessary to sustain the rapid heat loss through the thin ice over the poles on Enceladus, motivated by the fact that the tidal dissipation in the ice may be insignificant \citep{Cadek-Soucek-Behounkova-et-al-2019:long}. According to our results, such a polar-amplified ocean heat flux may be hard to achieve on Enceladus, given its strong ice topography.
In fact, as can be seen from the blue and green curves in Fig.~\ref{fig:heat-transport}c-e, ice shell over the poles is losing heat toward the ocean in both shell-heating and core-heating scenarios. If the ice does not produce any heat as assumed in the core-heating scenario, the polar ice shell will freeze due to the heat loss through the ice and toward the ocean. Meanwhile, ice flow will transport ice poleward. Both freezing and ice flow will remove the poleward thinning ice topography over time.

\section{Sensitivity tests and remarks}
\label{sec:sensitivity-test}
\textbf{Convection.} As mentioned in section~\ref{sec:model-configuration}, our limited model resolution may lead to underestimation of the Nusselt number and thereby overestimation of $\Delta_vT$. This bias may exist but does not seem to be very pronounced, because if we substitute the $a=250$~km model setups into Eq.~\eqref{eq:deltavT-Nu}, we will get a $\Delta_vT$ of 3~mK, 5~mK and 9~mK at $Q_0=Q_c/3,\ Q_c$ and $5Q_c$, respectively. These are roughly consistent with our numerical solutions shown in Fig.~C1, indicating the convective heat transport may be somewhat captured. Besides, the convective dynamics presented in Fig.~\ref{fig:solution-a1000}e5 and in Fig.C8-10 are actually qualitatively similar to the solution obtained in DNS \citep[e.g.,][]{Soderlund-2019:ocean}.

However, study the potential impact of such bias is still necessary. We thus conducted an extra set of experiments with strong convective parameterization. In these experiments, we let the water column mix at a rate that equals two times what is predicted by the non-diffusive scaling (Eq.~\ref{eq:kappa-conv}, \citet{Gastine-Wicht-Aubert-2016:scaling}). As shown by Fig.~C4, the vertical temperature gradient is indeed decreased by the parameterization, but the horizontal heat deflection is hardly affected. This is also manifested by the overlap between the small triangular symbols (w/ convective parameterization) and the small circles (wo/ convective parameterization, default) in Fig.~\ref{fig:heat-transport}a. However, it should be noted that the convection parameterization adopted here only mixes tracers along the direction of gravity, which applies the same mixing rate for the entire domain, regardless of the local supercriticality, and does not account for the fact that the convection-induced mixing should be aligned with the rotating axis rather than the gravity. More sophisticated parameterzation based on high-resolution local simulations needs to be developed to better understand the interaction between baroclinic eddies and convection on icy moons.

\textbf{Boundary exchange coefficient.} Sensitivity to ocean-ice heat exchange coefficient $\gamma_T$ is also tested for all the strongly forced experiments. The heat deflections by ice topography in the experiments with $50$ times larger $\gamma_T$ are shown by square symbols in Fig.~\ref{fig:heat-transport}a. Compared to the default experiments (shown by circles), the heat deflection is increased. This is expected because more efficient ocean-ice heat exchange will decrease the temperature gradient across the boundary layer at water-ice interface, especially in low latitudes where a large amount of heat needs to be delivered to the ice. This will increase the meridional temperature gradient $\Delta_hT$, as can be seen by comparing row-a of Fig.C5-7 on one hand against Fig.~\ref{fig:solution-a1000} and Fig.C1-2 on the other. The increased $\Delta_hT$ then enhances the meridional OHT, as indicated by the upward shift of square symbols in Fig.~\ref{fig:heat-transport}a relative to the circles and by the more equatorward-amplified surface heat flux pattern shown by the dashed curves in Fig.~\ref{fig:heat-transport}c-e. 
In fact, if we subtract $2Q_c/(\gamma_T\rho C_p)$ from $\Delta_hT$ in Eq.~(\ref{eq:Qc}), we can solve for a $\gamma_T$-modified $\tilde{Q_c}$ from
  \begin{equation}
    \label{eq:Qc-gamma-modified}
    \tilde{Q_c}=\frac{k C_{p} \rho_{0}}{\tilde{a}^3 f^{2}}(\alpha g)^{\frac{3}{2}} D^{\frac{5}{2}}\left(\Delta_h T -\frac{2\tilde{Q_c}}{\gamma_T\rho C_p}\right)^{\frac{5}{2}}.
  \end{equation}
  This new $\tilde{Q_c}$ can be solved numerically and will be smaller than the $Q_c$ given by Eq.~(\ref{eq:Qc}), indicating again that the heat deflection by ice topography will be weakened by inefficient heat exchange between ice and ocean (i.e., a small $\gamma_T$). Shown in Fig.~\ref{fig:heat-transport}f is $\mathcal{H}_{\mathrm{ocn}}/\tilde{Q_c}$ plotted as a function of $Q_0/\tilde{Q_c}$ in log-log scale. Normalizing $\mathcal{H}_{\mathrm{ocn}}$ and $Q_0$ with the $\gamma_T$-modified $\tilde{Q_c}$ instead of the $Q_c$ given by Eq.~(\ref{eq:Qc}) increases the correlation coefficient slightly from $0.92$ to $0.94$.

  \textbf{Latitudinal range that can receive bottom heating.}
  It can also be seen from Fig.~\ref{fig:heat-transport}c-e that, in the nearly critical or supercritical scenarios, the heat flux from the ocean remains rather evenly distributed in low latitudes until it suddenly drops. This feature arises because the high-latitude ice shell can hardly receive any heat unless the isopycnal initiated from the ice at that latitude reaches the convective region at the bottom, where continuous heat supply is available (see schematics Fig.~\ref{fig:schematics}c). Otherwise, the stable stratification under the ice would not only prevent heat from reaching the ice shell but also cause heat diffusion away from the ice. The stably stratified layer under the polar ice can be clearly seen in the high $\gamma_T$ experiments (Fig.C5-7), but is less visible in the control experiments (Fig.~\ref{fig:solution-a1000} and Fig.C1-2). This is because the inefficient ocean-ice heat exchange (low $\gamma_T$) in the control experiments causes heat to accumulate in the ocean, and this accumulation of heat makes the sub-ice ocean much warmer than the freezing point so that the prescribed amount of heat can be transmitted to the ice. This raises the temperature of entire ocean, making the warm freezing point under polar ice shell seems less warm in comparison. As a result, the polar stratification zone becomes less visible. In very high latitudes, there are places where freezing point is indeed higher than the water underneath, but stratified layer tends to be much thinner (Fig.~\ref{fig:solution-a1000} and Fig.C1-2). Due to this reason, the high $\gamma_T$ experiments (Fig.C5-7) matches the physical picture we sketched in Fig.~\ref{fig:schematics}c better.

  Suppose cutoff of ocean heat flux happens around latitude $\theta_c$, where the freezing point under the ice $T_{c}$ is warmer than the equatorial freezing point $T_{eq}$ by $T_{c}-T_{eq}=b_0\rho_ig(1.5 \Delta H \sin^2\theta_c)$ (ice topography follows 2nd order Legendre polynomial of $\sin\theta_c$). If we define the cutoff threshold to be when the heat flux from the ocean drops to $1/4$ of its peak value, the criterion for the isopycnal that initiates at $\theta_c$ under the ice to touch the seafloor can be determined by letting upward heat transport per area $\mathcal{F}_v/(\pi \tilde{a}^2)$ equals $1/4$ of the bottom flux $Q_0$. Substituting Eq.~\eqref{eq:critical-F}, we get
  \begin{equation}
    \label{eq:Tc-Teq}
    T_{c}-T_{eq}-\left(1-\frac{1}{4}\right)\frac{Q}{\gamma_T\rho C_p}=\Delta_hT=\left(\frac{Q \tilde{a}^3f^2}{8k\rho C_p}\right)^{\frac{2}{5}}(\alpha g)^{-\frac{3}{5}}D^{-1},
  \end{equation}
  where the last term in the first equality accounts for the temperature jump across the ocean boundary layer under the ice due to the limited heat exchange coefficient $\gamma_T$. Solving the above equation would allow us to predict $ T_{c}-T_{eq}$ and thereby the transitional latitude $\theta_c$. As can be seen from Fig.~\ref{fig:heat-transport}(g,h), our prediction matches the diagnosed $ T_{c}-T_{eq}$ and $\theta_c$ reasonably well. $\theta_c$ is diagnosed from simulations by finding the latitude where $\mathcal{H}_{\mathrm{surf}}$ decreases to $1/4$ of its peak value in Fig.~\ref{fig:heat-transport}c-e.

  \textbf{Salinity factor.} In this work, we have focused on the ocean dynamics driven by temperature variations, however, salinity factor could be important especially on smaller icy moons for reasons discussed in \citet{Kang-Jansen-2022:icy, Kang-2022:different}. We thus did sensitivity tests for the core-heating and shell-heating cases with $a=250$~km, using the full nonlinear ``MDJWF'' equation of state \citep{McDougall-Jackett-Wright-et-al-2003:accurate}, which automatically accounts for both the temperature and salinity-induced density anomalies. The solutions and heat transport behaviors remain qualitatively similar to the temperature-only experiments, as shown by Fig.~C3 and the cross signs in Fig.~\ref{fig:heat-transport}b. With an even larger bottom heating (e.g., $Q_c$ or $5Q_c$), the ocean transport becomes even more efficient. That necessarily means that the salinity gradient between the equator and the poles will be even less given that the freezing/melting rate and salinity flux is determined by the speed of the ice flow (Eq.~B8) assuming equilibrium state.

  We also note that, if the ocean salinity is sufficient low ($<22$~psu for Enceladus), the density anomalies associated with salinity to cancel out with that associated with temperature, and the direction of the ocean residual circulation could reverse, i.e., the dense warm water formed under the polar ice shell sinks \citep{Kang-Mittal-Bire-et-al-2022:how}. Because the elevation of dense water formation (under the thin polar ice shell) is higher than the elevation of buoyant water formation (under the thick equatorial ice shell), extra energy is injected to the ocean \citep{Jansen-Kang-Kite-2022:energetics}. This energy can potentially drive stronger ocean dynamics. How the bottom heating interacts with this reversed meridional density gradient at the surface is yet to be explored.

  \section{Implications for Enceladus and other icy satellites and discussions}
  \label{sec:implications}
  Now that we have demonstrated that 1) the amount of bottom heat flux required to overcome the stratification under the thin ice can be predicted using $Q_c$ from \eqref{eq:Qc} and 2) a bottom heat flux of even $5Q_c$ is insufficient to preserve its pattern as being transported to the ice, we can use these to put an upper bound on the Enceladus bottom heating and on the ice thickness variation that can be induced by non-uniform bottom heating on an arbitrary icy moon.

  Since, on Enceladus, $Q_c=2$~W/m$^2$ if thermal expansivity $\alpha=10^{-4}$/K and $Q_c=63$~mW/m$^2$ if $\alpha=10^{-5}$/K \footnote{$|\alpha|<10^{-5}$/K requires the mean salinity to be between $18$ and $26$~psu at Enceladus pressure.}, a bulk bottom heat flux in order of at least $5Q_c=10$~W/m$^2$ and $0.3$~W/m$^2$, respectively, is needed in order to deliver any significant amount of heat to the polar ice shell, and even stronger heat flux is needed to preserve the bottom heating pattern in upward heat transport. This is unlikely to be achievable given that the global-averaged heat production rate is merely $40$~mW/m$^2$. Locally focused bottom heating as proposed by \citet{Choblet-Tobie-Sotin-et-al-2017:powering} may not help either, because baroclinic instability induced by the temperature contrast between the plume water and the ambient water has been shown to be able to efficiently homogenize the heat flux laterally in the scale of tens of kilometers \citep{Kang-Marshall-Mittal-et-al-2022:ocean}. Therefore, the observed ice geometry on Enceladus \citep{Hemingway-Mittal-2019:enceladuss, McKinnon-Schenk-2021:new} is unlikely to be sustainable by heating generated in the silicate core alone. The same conclusion has been derived by a series of previous works by examining the ice shell heat budget \citep{Kang-Mittal-Bire-et-al-2022:how, Kang-Marshall-Mittal-et-al-2022:ocean} and the feedback of ocean heat transport on hemispheric asymmetry of Enceladus' ice shell \citep{Kang-Bire-Marshall-2022:role}.

  Furthermore, unless $|\alpha|\lesssim 10^{-5}$/K, the bulk bottom heat flux is unlikely to exceed $Q_c$, so it is likely that almost all of the bottom heating is deflected toward the equator (see red curves in Fig.~\ref{fig:heat-transport}c-e). Considering that new ice formation must occur continuously at the equator to counterbalance the poleward mass transport caused by ice flow \citep{Ashkenazy-Sayag-Tziperman-2018:dynamics, Kang-Flierl-2020:spontaneous}, the total power released from the seafloor should not surpass the conductive heat loss rate through the equatorial ice shell (30S-30N), which is approximately 3~GW. Even with $|\alpha|\lesssim 10^{-5}$/K, $5Q_c$ is still expected to exceed the bulk bottom heat flux, resulting in over half of the bottom heating being deposited onto the equatorial ice shell (see black curves in Fig.~\ref{fig:heat-transport}c-e). For the ocean to freeze over the equator, the total power released from the silicate core must not exceed 6~GW. 

  On larger icy moons, such as Europa, Titan, Ganymede and Callisto, even moderate ice topography can induce strong horizontal temperature contrast under the ice, $\Delta_hT$, because of the stronger gravity (Eq.~\ref{eq:deltahT}). In the meanwhile, the vertical temperature gradient $\Delta_vT$ decreases as a result of the more efficient convective mixing (Eq.~\ref{eq:deltavT-Nu}). This necessarily makes it easier for $\Delta_hT$ to dominate $\Delta_vT$. If the heterogeneity or entirety of insolation can be neglected (see insolation-driven ocean dynamics in \citet{Ashkenazy-Tziperman-2021:dynamic}), the largest possible ice thickness contrast can be sustained by inhomogeneous bottom heating \citep{Liao-Nimmo-Neufeld-2020:heat, Choblet-Tobie-Sotin-et-al-2017:powering} can be estimated by equating $5Q_c$ with $\mathcal{H}_{\mathrm{cond}}$ (Eq.~B10),
  \begin{equation}
    \label{eq:max-dH}
    (\Delta H)_{\mathrm{max}} =(b_0\rho_i gD)^{-1}\left(\frac{\kappa_{0}\tilde{a}^3 f^{2}}{5k C_{p} \rho_{0}H_0} \ln \left(T_{f}/T_{s}\right)\right)^{\frac{2}{5}}(\alpha g)^{-\frac{3}{5}},
  \end{equation}
  where $H_0$ is the mean thickness of ice, $T_s$ is the surface temperature and $T_f\sim 273$K is the freezing point of water. Substituting Europa parameters ($T_s=110$K, $\alpha=2\times 10^{-4}$/K, $H_0=15$~km, $D=85$~km), we get an $(\Delta H)_{\mathrm{max}}$ of merely $88$~m. This indicates that inhomogeneous bottom heating alone can hardly sustain any ice thickness variation on large icy satellites without inhomogeneity in ice heat production.

  Icy satellites are not the only planetary bodies that are simultaneously forced by internal heating from below and laterally varying surface temperature. Gas/ice giants inside our solar system, namely, Jupiter, Saturn, Uranus and Neptune, too are forced by internal and surface heat forcings, except that the internal heat originates from the initial gravitational accretion \cite{Gierasch-Ingersoll-Banfield-et-al-2000:observation, Guillot-2005:interiors} and the surface is forced by solar radiation \cite{Levine-Kraemer-Kuhn-1977:solar}. Except Uranus, which has an obliquity close to 90 degree, the other three planets receive very little solar energy over the polar regions, which would allow the internal-heating-driven convective plumes to penetrate, analogous to the thick ice regions on icy satellites. On the other hand, the low latitude upper atmospheres on Jupiter, Saturn and Neptune are very likely to be stably stratified due to the solar heating at the surface. The framework proposed in our work may be modified to predict the stratification boundary on these gas giants and ice giants.

  Lastly, it is important to mention that, when $\Delta_hT$ dominates over $\Delta_vT$, which is true for Enceladus, the ocean just under the ice will be stably-stratified. This stratification will largely impede the tracer transport, forcing the water within the convective plume to mix with the surrounding water, as pointed out by \citet{Zeng-Jansen-2021:ocean} and \citet{Kang-Marshall-Mittal-et-al-2022:ocean}. Consequently, biosignatures, which are likely to be concentrated within the convective plume, will be substantially diluted before being ejected.

\section*{Acknowledgements}
This work is carried out in the Department of Earth, Atmospheric and Planetary Science (EAPS) MIT using svante cluster (a total 6,000,000 core-hours is devoted to this work). WK acknowledges support by the 2023 research committee grant from MIT and helpful interaction with the Exploring Ocean World (ExOW) group funded by NASA Astrobiology Grant 80NSSC19K1427. All data can be reproduced following the model description in the appendix B, and code is available upon reasonable request.

\section*{Data Availability}

 The supporting information provides detailed description of our model setup. We would like to provide more code and data upon reasonable request.



\bibliographystyle{mnras}
\bibliography{export} 



\appendix
\section{Derivation of horizontal and vertical ocean heat transport $\mathcal{F}_h$, $\mathcal{F}_v$.}
\label{sec:derivation-OHT}

 In a stratified fast-rotating system forced with horizontal temperature gradient, both overturning circulation and baroclinic eddies can transport heat downgradient. A single-cell overturning circulation can be sustained only when both top and bottom surfaces are sufficiently frictious, and the resultant circulation flows along the direction of rotation in the interior, the circulation can be closed at the two boundaries \cite{Kang-Jansen-2022:icy}. This circulation transports tracers (including heat and salinity) downgradient meridonally while creating stratification in the fluid. Typically, after the adjustment by single-cell overturning circulation, the system is still subject to baroclinic instability, which further extracts energy from the gravity potential of the water and converts that to eddy kinetic energy. 
 Although the fluid patterns associated with baroclinic eddies are totally different from that associated with an overturning cell, their effect on heat transport are very similar. In earth atmosphere, forced by the poleward cooling trend, baroclinic eddies transport heat poleward (downgradient) to reduce the meridional temperature contrast, lower the center of the mass of the fluid to create stratification, transport prograde zonal momentum upward to form the vertical shear required by the thermal-wind relationship, and meanwhile, converge zonal momentum to regions where eddies are generated to form jets \cite{Vallis-2006:atmospheric}.

Here, we ignore the heat transport induced by the single-cell overturning circulation, given its likely small magnitude \cite{Kang-2022:different}, and focus on heat transport by baroclinic eddies. In an ocean with a meridional temperature gradient of $T_y=\Delta_hT/\tilde{a}$, eddy motions will transport heat downgradient. The total heat transport $\mathcal{F}_h$ as a function of eddy diffusivity $\kappa_e$ following the mixed-length theory \cite{Vallis-2006:atmospheric},
  \begin{equation}
    \label{eq:Fh1}
    \mathcal{F}_h=(2\pi \tilde{a}d)C_p \rho_0\overline{v'T'}=(2\pi \tilde{a}d)C_p \rho_0 \kappa_e T_y \sim (2\pi \tilde{a}) C_p\rho_0  \kappa_es\Delta_vT,
  \end{equation}
  where $s\equiv T_y/T_z=(\Delta_hT/\tilde{a})/(\Delta_vT/d)$ denotes the slope of the isopycnals, $d$ is the penetration depth of the surface density anomaly, $\pi a$ is the circumference of the zonal circle in mid-latitudes, and $\Delta_hT$ and $\Delta_vT$ are the horizontal and vertical temperature contrast across the domain. 

  Since in stratified fluid, fluid parcels are confined to move along isopycnals \citep{Vallis-2006:atmospheric}, the heat flux must be aligned with isopycnals too \footnote{Perpendicular to the isopycnals, because no fluid motion is permitted, heat flux should also vanish.}, as sketched in Fig.~\ref{fig:schematics}a,c and demonstrated in Fig.~\ref{fig:schematics}d. The total vertical heat transport $\mathcal{F}_v$, therefore, can be written as
  \begin{equation}
    \label{eq:Fv1}
    \mathcal{F}_v=\mathcal{F}_h\frac{\Delta_hT}{\Delta_vT}\sim (2\pi \tilde{a}) C_p\rho_0  \kappa_es\Delta_hT.
  \end{equation}
  
  In the literature, a so-called residual circulation $\Psi^\dagger$ is typically introduced to represent the eddy mixing with an equivalent overturning circulation. It takes the form:
  \begin{equation}
    \label{eq:Psi-residual}
    \Psi^\dagger\sim (2\pi \tilde{a}) \kappa_e.
  \end{equation}
  With $\Psi^\dagger$, the heat transports can be written as
  \begin{equation}
    \label{eq:Fv-Fh-Psi}
    \mathcal{F}_h=\Psi^\dagger\Delta_vT,\ \mathcal{F}_v=\Psi^\dagger\Delta_hT.
  \end{equation}

  The eddy diffusivity $\kappa_e$ can be expressed as the product of eddy characteristic size $L_e$ and the eddy characteristic speed $V_e$ times $k=0.25$ following the mixing length theory \cite{Vallis-2006:atmospheric}.

 \textbf{Ocean heat transport when isopycnals do not touch the seafloor -- derivation of Eq~\eqref{eq:critical-F}.} 
  When the isopycnals originated from the ice do not or marginally touch the ground, we have $\Delta_vT=\Delta_hT$, $L_e$ can be approximated by the deformation radius $L_d\equiv \frac{Nd}{f}$, and $V_e$ can be approximated by the thermal wind speed (bottom minus top zonal speed) $U$. This gives
  \begin{equation}
    V_e\sim  U\sim \frac{\alpha_T g\Delta_h T d}{2\Omega \tilde{a}},\ 
    L_e\sim L_d\sim \frac{\sqrt{\alpha_T g\Delta_v Td}}{2\Omega} \label{eq:Ve-Le-subcritical}
  \end{equation}
  Substituting Eq~(\ref{eq:Ve-Le-subcritical}) into Eq~\eqref{eq:Fh1} and Eq.~\eqref{eq:Fv1}, we get
  \begin{equation}
    \label{eq:Fh-subcritical}
    \mathcal{F}_h=\mathcal{F}_v= \frac{2 \pi k C_{p} \rho_{0}}{ f^{2}}(\alpha g \tilde{a})^{\frac{3}{2}} (\Delta_h T s)^{\frac{5}{2}},
  \end{equation}
  which is the first equality in Eq~\eqref{eq:critical-F}. Note that because $\Delta_hT=\Delta_vT$ when isopycnals do not touch the seafloor, the vertical heat transport $\mathcal{F}_v$ equals the horizontal heat transport $\mathcal{F}_h$. 

  \textbf{Ocean heat transport when isopycnals touch both ice shell and seafloor.}
  It should be noted that $L_e\sim L_d$, $V_e\sim U$ only holds when $\Delta_vT=\Delta_hT$. The more general scaling should be $V_e/U\sim L_e/L_d\sim \Delta_hT/\Delta_vT\sim (\tilde{a}s/d)$ as shown by \citet{Held-Larichev-1996:scaling}. Eq~\eqref{eq:Ve-Le-subcritical} can be recovered with $d=\tilde{a}s$. In contrast, when isopycnals touch the seafloor, $d$ is confined by the ocean depth $D$ and the factor $(\tilde{a}s/d)=\tilde{a}s/D=\Delta_hT/\Delta_vT>1$. This leads to
    \begin{equation}
    V_e\sim \frac{\tilde{a}s}{D} U\sim \frac{\alpha_T g\Delta_h T s}{2\Omega },\ 
    L_e\sim \frac{\tilde{a}s}{D} L_d\sim \frac{\sqrt{\alpha_T g\Delta_hT\tilde{a}s}}{2\Omega}, \label{eq:Ve-Le-supercritical}
  \end{equation}
  This scaling can be obtained following \citet{Held-Larichev-1996:scaling}, when the following assumptions are made: 1) the potential energy embedded in the meridional density gradient is converted to kinetic energy at the scale of deformation radius $L_d$, 2) the baroclinic kinetic energy cascades to barotropic kinetic energy through the interaction between two baroclinic modes to form 2D geostrophic turbulence, and 3) energy of 2D geostrophic turbulence cascades to large scales until halted at the Rhines scale. See \citet{Vallis-2006:atmospheric} chapter~9 for 2D turbulence energy cascade.
  With Eq~\eqref{eq:Ve-Le-supercritical}, the meridional heat flux should instead scales as follows,
  \begin{equation}
    \label{eq:F-supercritical}
    \mathcal{F}_h=\frac{2\pi k C_p\rho_0}{f^2}(\alpha_T g )^{\frac{3}{2}}\tilde{a}^{1/2}D(\Delta_h T )^{\frac{5}{2}}s^{\frac{3}{2}}.
  \end{equation}
  After isopycnals touch the seafloor, their further steepening would make parts of isopycnals sink beneath the seafloor, preventing the meridional heat transport to reach its full potential. This is manifested by a weaker sensitivity to $s$ in Eq~\eqref{eq:F-supercritical} than Eq~\eqref{eq:critical-F}.

  The vertical heat transport $\mathcal{F}_v$ in this case is greater than instead of equal to the horizontal heat transport $\mathcal{F}_h$,
  \begin{equation}
    \label{eq:Fv-supercritical}
    \mathcal{F}_v=(2\pi \tilde{a}) C_p\rho_0  \kappa_es\Delta_hT=\mathcal{F}_h\frac{\tilde{a}s}{D}=\frac{2\pi k C_p\rho_0}{f^2}(\alpha_T g )^{\frac{3}{2}}\tilde{a}^{3/2}(\Delta_h T s)^{\frac{5}{2}} > \mathcal{F}_h.
  \end{equation}
 Mathematically, $\mathcal{F}_v$ above takes the same form as the $\mathcal{F}_h,\mathcal{F}_v$ scaling for cases where isopycnals do not touch the seafloor.

\textbf{Energetics of baroclinic eddies.}
 From the energetic point of view, the baroclinic eddies convert potential energy to kinetic energy at a rate of
  \begin{equation}
    \label{eq:energy-conversion-baroclinic}
    \epsilon_{\mathrm{bc}}=\kappa_e U^2/L_d^2=\frac{k}{f^2\tilde{a}^3}(\alpha g \Delta_hT d)^{5/2},
  \end{equation}
  per kilogram of water, under quasi-geostrophy approximation (see \cite{Held-Larichev-1996:scaling}). The consumed potential energy will be replenished by bottom heating at exactly the same rate
  \begin{equation}
    \label{eq:energy-conversion-convection}
    \epsilon_{\mathrm{conv}}=\frac{\alpha g Q}{C_p\rho},
  \end{equation}
  as long as the mean bottom heat flux $Q$ equals the mean upward heat transport by baroclinic eddies
  \begin{equation}
    \label{eq:vertical-heat-fluxes}
    Q=\mathcal{F}/(2\pi \tilde{a}^2)=\frac{k\rho C_p}{f^2\tilde{a}^3}(\alpha g)^{3/2} (\Delta_hT d)^{5/2}.
  \end{equation}
  This heat flux balance should hold in order for the temperature profile to be in steady state.

  \textbf{Derivation of ocean heat transports without bottom heating.}
  In absence of bottom heating, the potential energy consumed by baroclinic eddies is replenished by downward diffusion of buoyancy \citep{Jansen-Kang-Kite-2022:energetics}. In an equilibrium state, the upward heat flux by baroclinic eddies need to balance the downward heat diffusion.
  \begin{equation}
    \label{eq:vertical-heat-balance-diffusion}
\mathcal{F}_v=\kappa_v\frac{\Delta_vT}{d} (2\pi a^2)
  \end{equation}
  Substituting the first equality of Eq~\eqref{eq:critical-F} and $\Delta_hT=\Delta_vT$ into the above equation would allow us to solve for $d$, which then can be substituted into Eq~\eqref{eq:critical-F} to obtain
  \begin{equation}
    \label{eq:OHT-scaling-shellheating}
    \mathcal{F}_v=\mathcal{F}_h=2\pi \rho C_p \Delta_h T \left(\frac{k}{af^2}\right)^{2/7}(\alpha g\Delta_h T)^{3/7}(\kappa_va^2)^{5/7}.
  \end{equation}
  This is identical to what is presented in \citet{Kang-2022:different}.

  \section{Description of the Ocean Model}
\label{sec:model-description}
  
  Our simulations are carried out using the Massachusetts Institute of Technology OGCM (MITgcm, \cite{MITgcm-group-2010:mitgcm, Marshall-Adcroft-Hill-et-al-1997:finite}) configured for application to icy moons. 
  
  The model integrates the non-hydrostatic primitive equations for an incompressible fluid in height coordinates, including a full treatment of the Coriolis force in a deep fluid, as described in \cite{MITgcm-group-2010:mitgcm, Marshall-Adcroft-Hill-et-al-1997:finite}. Such terms are typically neglected when simulating Earth's ocean because the ratio between the fluid depth and horizontal scale is small. Instead, when the moon size is order hundreds of kilometers like Enceladus, the aspect ratio is order $0.1$ and so not negligibly small. The size of each grid cell shrinks with depth due to spherical geometry and is accounted for by switching on the ``deepAtmosphere'' option of MITgcm. Also, the gravity will vary with depth as well. This is accounted for using the following profile of gravity.
s  \begin{equation}
    \label{eq:g-z}
    g(z)=\frac{4\pi G\left[\rho_{\mathrm{core}}(a-D_0-H_0)^3+\rho_{\mathrm{out}}((a-z)^3-(a-D_0-H_0)^3)\right]}{3(a-z)^2}.
  \end{equation}
  In the above equation, $G=6.67\times10^{-11}$~N/m$^2$/kg$^2$ is the gravitational constant, $\rho_{\mathrm{core}}=2500$~kg/m$^3$ is the assumed core density and $\rho_{\mathrm{out}}=1000$~kg/m$^3$ is the density of the ocean/ice layer. $D_0$ and $H_0$ is the thickness of ocean and ice on global average.
  
  Since it takes several tens of thousands of years for our solutions to reach equilibrium, all of our experiments are first run under a zonally symmetric 2D configuration with a moderate resolution of $2$~degree ($8.7$~km) and 30 layers (each $\sim 2.5$~km) are used to keep the computational cost manageable. After equilibrium is reached, I interpolate the pick up files to generate initial conditions for the corresponding 3D simulation, which has a default horizontal resolution of 0.25$\times$0.25~degree and 70 unevenlly distributed vertical layers, whose thicknesses increase from 500~m to 2~km from top to bottom. Limited by our vertical resolution, we cannot probe the regime where isopycnal penetration depth $d$ is comparable to or below 2~km. This does not happen to any of our experiments, because the vertical diffusivity $\kappa_v$ is set to $10^{-3}$~m$^2$/s which guarantees that the penetration depth is in order of 10~km following the scaling laws by \citet{Kang-2022:different}, which is briefly reviewed in the appendix A. On top of that, bottom heating also deepens the isopycnals' penetration depth. By design, the supercritical cases ($Q_0=Q_c,\ 5Q_c$) have penetration depth greater than the ocean depth.

  \subsection{Diffusivity and Viscosity.}
  Vertical diffusivity affects the energetics of the ocean \cite{Young-2010:dynamic, Jansen-Kang-Kite-2022:energetics}. To account for the mixing of heat and salinity by unresolved turbulence, in our calculations, I set the explicit vertical diffusivity to $0.001$~m$^2$/s in both 2D and 3D simulations. This is estimated by substituting ocean dissipation rate by \cite{Rekier-Trinh-Triana-et-al-2019:internal} into the scaling reviewed by \cite{Wunsch-Ferrari-2004:vertical}. In all experiments, explicit horizontal diffusivity is set to be equal to the vertical diffusivity regardless of the resolution and the size of the icy moon.

  To keep the simulation stable, the explicit horizontal and vertical viscosity is set to 0.001~m$^2$/s and 0.03~m$^2$/s, respectively, and the widely-applied Smagorinsky viscosity scheme \cite{Smagorinsky-1963:general} is employed to represent the mixing in strongly sheared flows. Unlike the fixed viscosity scheme, Smagorinsky scheme determines the viscosity based on the resolved dynamics, and as a result, numerical noise will be damped while dynamics can be kept to a larger extent. The Smagorinsky viscosity constant is set to $3$ by default. As mentioned before, resolution in the x-direction is increased (decreased) by a factor of 1.4 under a higher (lower) rotation rate. In those experiments, horizontal viscosity is adjusted proportional to the x-grid width.

  \subsection{Equation of state and the freezing point of water}
  To make it easier to decipher the results, we adopt the ``LINEAR'' equation of state, where thermal expansion coefficient $\alpha$ and haline contraction coefficient $\beta$ are set as constants. Furthermore, since temperature-induced density variation is likely to dominate in the current configuration \cite{Kang-2022:different}, we set $\beta=0$ and focus on the temperature-forced circulation. The thermal expansivity $\alpha$ for each experiment is chosen to be that evaluated at a salinity of 60~psu and a depth of 24~km.
  
  The freezing point of water $T_f$ is assumed to depend on local pressure $P$ and salinity $S$ as follows,
  \begin{equation}
    \label{eq:freezing-point}
    T_f(S,P)=c_0+b_0P+a_0S,
  \end{equation}
where $a_0=-0.0575$~K/psu, $b_0=-7.61\times10^{-4}$~K/dbar and $c_0=0.0901$~degC. The pressure $P$ can be calculated using hydrostatic balance $P=\rho_igH$ ($\rho_i=917$~kg/m$^3$ is the density of the ice and $H$ is the ice thickness).
  
\begin{figure*}
    \centering \includegraphics[page=2,width=0.6\textwidth]{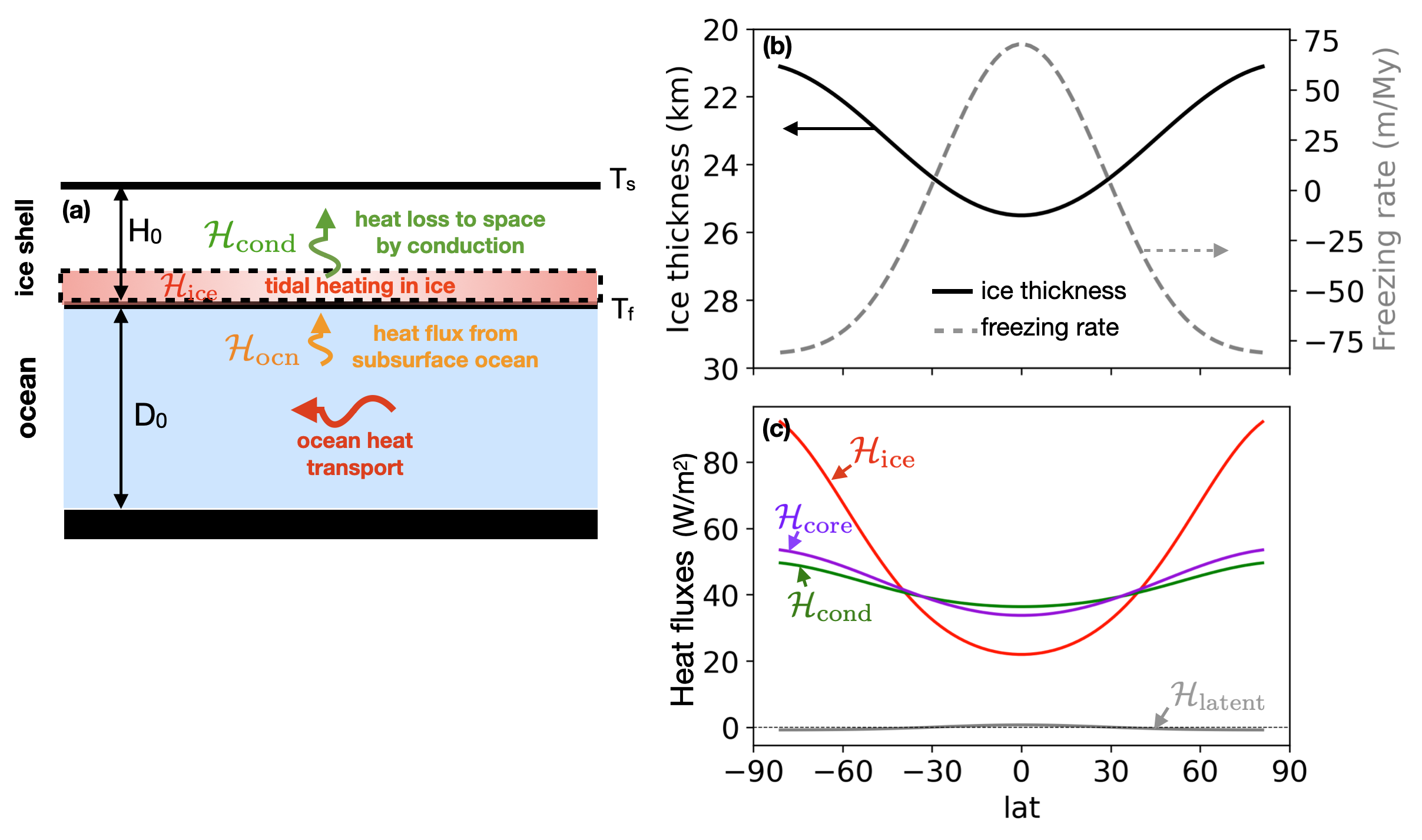}
    \caption{\small{Model configurations. Panel (a) defines the primary sources of heat and heat fluxes, which include: heating due to tidal dissipation in the ice $\mathcal{H}_{\mathrm{ice}}$, the heat flux from the ocean to the ice $\mathcal{H}_{\mathrm{ocn}}$ and the conductive heat loss to space $\mathcal{H}_{\mathrm{cond}}$. Panel (b) shows the ice shell thickness profile considered here using a black solid curve. The gray dashed curve shows the freezing (positive) and melting rate (negative) required to maintain a steady state based on an upside-down shallow ice flow model. As shown in section~\ref{sec:ice-flow}, the freezing/melting rate is inversely proportional to $a$. In this calculation, satellite radius $a$ is assumed to be 2500~km. Panel (c) shows the profiles of $\mathcal{H}_{\mathrm{ice}}$, $\mathcal{H}_{\mathrm{cond}}$ and $\mathcal{H}_{\mathrm{latent}}$.   }}
    \label{fig:configurations}
  \end{figure*}

  \subsection{Boundary conditions}
  \label{sec:boundary-conditions}

  The ocean is sandwiched between the silicate core and the ice shell. In this work, an equator-to-pole ice thickness gradient is prescribed such that the ice thickness profile follows
  \begin{equation}
    \label{eq:Hice}
    H(\phi)=H_0-H_2P_2(\sin\phi),
  \end{equation}
  where $H_0$ is the mean ice thickness, $P_2$ is the 2nd order Legendre polynomial, and $H_2$ is the amplitude of the ice thickness variation. $\phi$ denotes latitude. The thickness profile is shown by a solid curve in Fig.1b of the main text. Partial cells is switched on to better represent the ice topography: water is allowed to occupy a fraction of the height of a whole cell with an increment of 10\%. Interactions between the ice shell and the ocean is taken care of by a modified version of the MITgcm's ``shelfice'' module \cite{Losch-2008:modeling}.

  The ocean is forced by heat and salinity fluxes from the ice shell at the top and heat flux from the silicate core at the bottom. Here, two heating scenarios are considered: one with all heat production happening in the ice shell as in \cite{Kang-2022:different} and the other with all heat production happening in the silicate core. For both cases, the global heat budget is in balance and the tidal dissipation in the ocean is neglected following \cite{Chen-Nimmo-2011:obliquity, Beuthe-2016:crustal, Hay-Matsuyama-2019:nonlinear, Rekier-Trinh-Triana-et-al-2019:internal}.
 
 \underline{Tidal heating in the core}
 
   Following \cite{Beuthe-2019:enceladuss} and \cite{Choblet-Tobie-Sotin-et-al-2017:powering}, the core dissipation $\mathcal{H}_{\mathrm{core}}$ is set to peak at the two poles following Eq.60 in \cite{Beuthe-2019:enceladuss},
  \begin{equation}
    \label{eq:H-core}
    \mathcal{H}_{\mathrm{core}}(\phi)=\bar{\mathcal{H}}_{\mathrm{core}}\cdot(1.08449 + 0.252257 \cos(2\phi) + 0.00599489 \cos(4\phi)),
  \end{equation}
  where $\phi$ denotes latitude and $\bar{\mathcal{H}}_{\mathrm{core}}$ is the global mean heat flux from the bottom. Since the global surface area shrinks going downward due to the spherical geometry, a factor of $\left.(a-H)^2\right/(a-H-D)^2$ ($H$ is ice thickness, $D$ is ocean depth) needs to be considered when computing $\bar{\mathcal{H}}_{\mathrm{core}}$. The expression within the bracket is normalized for the globe, adjusted to take account of the fact that our model only covers 84S-84N. Using the above formula, the bottom heat flux is twice as strong over the poles than equator, as shown by the purple curve in Fig.1c of the main text. I note that the heating profile here is highly idealized and does not have the localized heating stripes seen in \cite{Choblet-Tobie-Sotin-et-al-2017:powering} which arise from the interaction between the porous core and the fluid in the gaps.

  \underline{Ice-ocean fluxes}

  The interaction between ocean and ice is simulated using MITgcm's ``shelf-ice'' package \cite{Losch-2008:modeling, Holland-Jenkins-1999:modeling} with some modifications. 

  At the water-ice interface, we consider the response of the ocean to a prescribed ice freezing rate while ignoring the possible response of the ice to the water-ice heat/salinity exchange. The freezing/melting induces a salinity/fresh water flux into the ocean (we assume the ice salinity to be zero); meanwhile, the ocean temperature at the upper boundary is relaxed to the local freezing point $T_f$ determined by the local salinity and pressure (Eq.~\ref{eq:freezing-point}).
  \begin{eqnarray}
    \frac{dS_{\mathrm{ocn-top}}}{dt}&=&\frac{qS_{\mathrm{ocn-top}}}{\delta z}\label{eq:S-tendency}\\
    \frac{dT_{\mathrm{ocn-top}}}{dt}&=&\frac{1}{\delta z}(\gamma_T-q)(T_f-T_{\mathrm{ocn-top}})\label{eq:T-tendency}
  \end{eqnarray}
  Here, $S_{\mathrm{ocn-top}}$ and $T_{\mathrm{ocn-top}}$ denote the upper boundary salinity and temperature, $\gamma_T=\gamma_S=10^{-5}$~m/s are the water-ice exchange coefficients for temperature and salinity, $\delta z=2$~km is the thickness of the water-ice ``boundary layer'' and $q$ is the freezing rate in m/s (note that $q$ is orders of magnitude smaller than $\gamma_T$). The ``boundary layer'' option is switched on to avoid possible numerical instabilities induced by an ocean layer which is too thin. In this work, the ice shell is assumed to be in mass balance, i.e., freezing/melting rate $q$ is prescribed to exactly compensate the ice thickness tendency induced by the ice flow (section~\ref{sec:ice-flow}).

  \underline{Friction drag}
  
  At the top and the bottom, tangential flow speed is relaxed back to zero at a rate of $\gamma_M=10^{-4}$m/s to mimic the friction drag.

  \subsection{Ice flow model}
  \label{sec:ice-flow}

  The prescribed freezing rate $q$ is computed using the divergence of the ice flow, assuming the ice sheet geometry is in equilibrium. Here, an upside-down land ice sheet model is used following \cite{Ashkenazy-Sayag-Tziperman-2018:dynamics}. The ice flows down its thickness gradient, driven by the pressure gradient induced by the spatial variation of the ice top surface, somewhat like a second order diffusive process. At the top, the speed of the ice flow is negligible because the upper part of the shell is so cold and hence rigid; at the bottom, the vertical shear of the ice flow speed vanishes, as required by the assumption of zero tangential stress there. This is the opposite to that assumed in the land ice sheet model. In rough outline, I calculate the ice flow using the expression below obtained through repeated vertical integration of the force balance equation (the primary balance is between the vertical flow shear and the pressure gradient force), using the aforementioned boundary conditions to arrive at the following formula for ice transport $\mathcal{Q}$,
\begin{equation}
  \mathcal{Q}(\phi)= \mathcal{Q}_0H^3(\partial_\phi H/a) \label{eq:ice-flow}
\end{equation}
where

\begin{equation}
\mathcal{Q}_0=\frac{2(\rho_0-\rho_i)g}{\eta_{m}(\rho_0/\rho_i)\log^3\left(T_f/T_s\right)}\int_{T_s}^{T_f}\int_{T_s}^{T(z)}\exp\left[-\frac{E_{a}}{R_{g} T_{f}}\left(\frac{T_{f}}{T'}-1\right)\right]\log(T')~\frac{dT'}{T'}~\frac{dT}{T}.\nonumber 
\end{equation}
Here, $\phi$ denotes latitude, $a$ and $g$ are the radius and surface gravity of the moon, $T_s$ and $T_f$ are the temperature at the ice surface and the water-ice interface (equal to local freezing point, Eq.~\ref{eq:freezing-point}), and $\rho_i=917$~kg/m$^3$ and $\rho_0$ are the ice density and the reference water density. $E_a=59.4$~kJ/mol is the activation energy for diffusion creep, $R_g=8.31$~J/K/mol is the gas constant and $\eta_{m}$ is the ice viscosity at the freezing point. The latter has considerable uncertainty ($10^{13}$-$10^{16}$~Pa$\cdot$s \cite{Tobie-Choblet-Sotin-2003:tidally}), and here $\eta_{m}$ is set to $10^{14}$~Pa$\cdot$s.

In steady state, the freezing rate $q$ must equal the divergence of the ice transport thus:
\begin{equation}
    q=-\frac{1}{a\cos\phi}\frac{\partial}{\partial \phi} (Q\cos\phi).
    \label{eq:freezing-rate}
\end{equation}
As shown by the dashed curve in Fig.1b of the main text, ice melts in high latitudes and forms in low latitudes at a rate of a few kilometers every million years. A more detailed description of the ice flow model can be found in \cite{Kang-Flierl-2020:spontaneous} and \cite{Ashkenazy-Sayag-Tziperman-2018:dynamics}. Freezing and melting leads to changes in local salinity and thereby a buoyancy flux.

\subsection{Heat budget}
If the ice shell is to be sustained, it needs to be in both mass and energy balance. The mass balance is assumed in our framework (Eq.\ref{eq:freezing-rate}) while the energy balance can be used to determine the equator-to-pole ice thickness gradient in equilibrium. The heat balance requires that the net heat surplus in the ice shell -- tidal dissipation in the ice $\mathcal{H}_{\mathrm{ice}}$, plus the heat flux from the ocean $\mathcal{H}_{\mathrm{ocn}}$ subtracting the conductive heat loss through the ice shell $\mathcal{H}_{\mathrm{cond}}$ -- equals the latent heat $\mathcal{H}_{\mathrm{latent}}=\rho_iL_fq$:
\begin{equation}
  \label{eq:heat-budget}
  \mathcal{H}_{\mathrm{ice}}+\mathcal{H}_{\mathrm{latent}}+\mathcal{H}_{\mathrm{ocn}}=\mathcal{H}_{\mathrm{cond}}.
\end{equation}

The main result of the paper is to express $\mathcal{H}_{\mathrm{ocn}}$ as a function of the equator-to-pole ice thickness gradient and other orbital and ocean parameters. $\mathcal{H}_{\mathrm{cond}}$ and $\mathcal{H}_{\mathrm{ice}}$ is estimated as follows.

  \underline{Diffusion of heat through the ice}
  
  Heat loss to space by heat conduction through the ice $\mathcal{H}_{\mathrm{cond}}$ is represented using a 1D vertical heat conduction model,
\begin{equation}
  \mathcal{H}_{\mathrm{cond}}=\frac{\kappa_{0}}{H} \ln \left(\frac{T_{f}}{T_{s}}\right),
  \label{eq:H-cond}
  \end{equation}
  where $H$ is the thickness of ice (solid curve in Fig.1b of the main text), the surface temperature is $T_s$ and the ice temperature at the water-ice interface is the local freezing point $T_f$ (Eq.~\ref{eq:freezing-point}).
  The surface temperature $T_s$ is set to the radiative equilibrium temperature, which can be computed given the incoming solar radiation and obliquity ($\delta=3^\circ$) and assuming an albedo of $0.81$. Typical heat losses averaged over the globe are $\mathcal{H}_{\mathrm{cond}}$= $50$~mW/m$^2$.

\underline{Model of tidal dissipation in the ice shell}

Icy moon's ice shell is periodically deformed by tidal forcing and the resulting strains in the ice sheet produce heat. I follow \cite{Beuthe-2019:enceladuss} to calculate the ice dissipation rate. Instead of repeating the whole derivation here, I only briefly summarize the procedure and present the final result. Unless otherwise stated, parameters are the same as assumed in \cite{Kang-Flierl-2020:spontaneous}.

Tidal dissipation consists of three components \cite{Beuthe-2019:enceladuss}: a membrane mode $\mathcal{H}_{\mathrm{ice}}^{\mathrm{mem}}$ due to the extension/compression and tangential shearing of the ice membrane, a mixed mode $\mathcal{H}_{\mathrm{ice}}^{mix}$ due to vertical shifting, and a bending mode $\mathcal{H}_{\mathrm{ice}}^{bend}$ induced by the vertical variation of compression/stretching. Following \cite{Beuthe-2019:enceladuss}, I first assume the ice sheet to be completely flat. By solving the force balance equation, I obtain the auxiliary stress function $F$, which represents the horizontal displacements, and the vertical displacement $w$. The dissipation rate $\mathcal{H}_{\mathrm{ice}}^{\mathrm{flat,x}}$ (where $x=\{\mathrm{mem},\mathrm{mix},\mathrm{bend}\}$ ) can then be written as a quadratic form of $F$ and $w$. In the calculation, the ice properties are derived assuming a globally-uniform surface temperature of 60K and a melting viscosity of $5\times10^{13}$~Pa$\cdot$s. 

Ice thickness variations are accounted for by multiplying the membrane mode dissipation $\mathcal{H}_{\mathrm{ice}}^{\mathrm{flat,mem}}$, by a factor that depends on ice thickness. The membrane mode is the only mode which is amplified in thin ice regions (see \cite{Beuthe-2019:enceladuss}). This results in the expression:
\begin{equation}
  \label{eq:H-tide}
  \mathcal{H}_{\mathrm{ice}}=(H/H_0)^{p_\alpha}\mathcal{H}_{\mathrm{ice}}^{\mathrm{flat,mem}}+\mathcal{H}_{\mathrm{ice}}^{\mathrm{flat,mix}}+\mathcal{H}_{\mathrm{ice}}^{\mathrm{flat,bend}},
\end{equation}
where $H$ is the prescribed thickness of the ice shell as a function of latitude and $H_0$ is the global mean of $H$. Since thin ice regions deform more easily and produce more heat, $p_\alpha$ is negative. Because more heat is produced in the ice shell, the overall ice temperature rises, which, in turn, further increases the mobility of the ice and leads to more heat production (the rheology feedback).


The tidal heating profile corresponding to $p_\alpha=-1.5$ is the red solid curve plotted in Fig.1c of the main text.

\begin{table*}[hptb!]
  
  \centering
\small
  \begin{tabular}{lll}
    Symbol & Name & Definition/Value\\
    \hline
    \multicolumn{3}{c}{Physical constants}\\
    \hline
    $L_f$ & fusion energy of ice & 334000~J/kg\\
    $C_p$ & heat capacity of water & 4000~J/kg/K\\
    $T_f(S,P)$ & freezing point & Eq.\ref{eq:freezing-point}\\
    $\rho_i$ & density of ice & 917~kg/m$^3$ \\
    $\rho_w$ & density of the ocean & 'LINEAR' equation of state \\
    $\kappa_0$ & conductivity coeff. of ice & 651~W/m\\
    $p_\alpha$ & ice dissipation amplification factor & -1.5 \\
    $\eta_{m}$ & ice viscosity at freezing point & 10$^{14}$~Ps$\cdot$s\\
    \hline
    \multicolumn{3}{c}{Default model setup}\\
    \hline
    $a$ & radius & 250, 1000, 2500~km\\
    $\alpha$ & thermal expansion coeff. &  $0.845, 1.05, 1.46\times 10^{-4}$/K for 3 different radii \\
    $\beta$ & saline contraction coeff. &  0\\
    $g_0$ & surface gravity & Eq.~\eqref{eq:g-z}\\
    $\delta$ & obliquity & 3.1$^\circ$\\
    $H_0$ & global mean ice thickness & 24~km  \\
    $H_2$ & equator-to-pole ice thickness variation & 3~km\\
    $H$ & ice shell thickness & Eq.\ref{eq:Hice}\\
    $D$ & global mean ocean depth& 52~km \\
    $\Omega$ & rotation rate & 2.05$\times$10$^{-5}$~s$^{-1}$ (3.5~day period)\\
    $\bar{T_s}$ & mean surface temperature& 62K\\
    $S_0$ & mean ocean salinity & 60~psu\\
    $P_0$ & reference pressure & $\rho_ig_0H_0$ \\
    $T_0$ & reference temperature & $T_f(S_0,P_0)$ \\
    $\nu_h$ & horizontal viscosity & 0.001~m$^2$/s (3D)\\
    $\nu_v$ & vertical viscosity & 0.03~m$^2$/s (3D)\\
    $\nu_{\mathrm{smag}}$ & Smagorinsky viscosity & 3 \\
    $\kappa_h,\ \kappa_v$ & horizontal/vertical diffusivity & 0.001~m$^2$/s\\
    $(\gamma_T,\ \gamma_S,\ \gamma_M)$ & water-ice exchange coeff. for T, S \& momentum & (10$^{-5}$, 10$^{-5}$, 10$^{-4}$)~m/s\\
    $\mathcal{H}_{\mathrm{cond}}$ & conductive heat loss through ice & Eq.\ref{eq:H-cond}\\
    $\mathcal{H}_{\mathrm{ice}}$ & tidal heating produced in the ice & Eq.\ref{eq:H-tide} \\
    \hline
     \end{tabular}
  \caption{Model parameters used in the ocean general circulation model. }
  \label{tab:parameters}
  
\end{table*}

\section{Solutions for larger icy satellites.}
\label{sec:exp-large-icy-moons}
\begin{figure*}
    \centering \includegraphics[page=6,width=\textwidth]{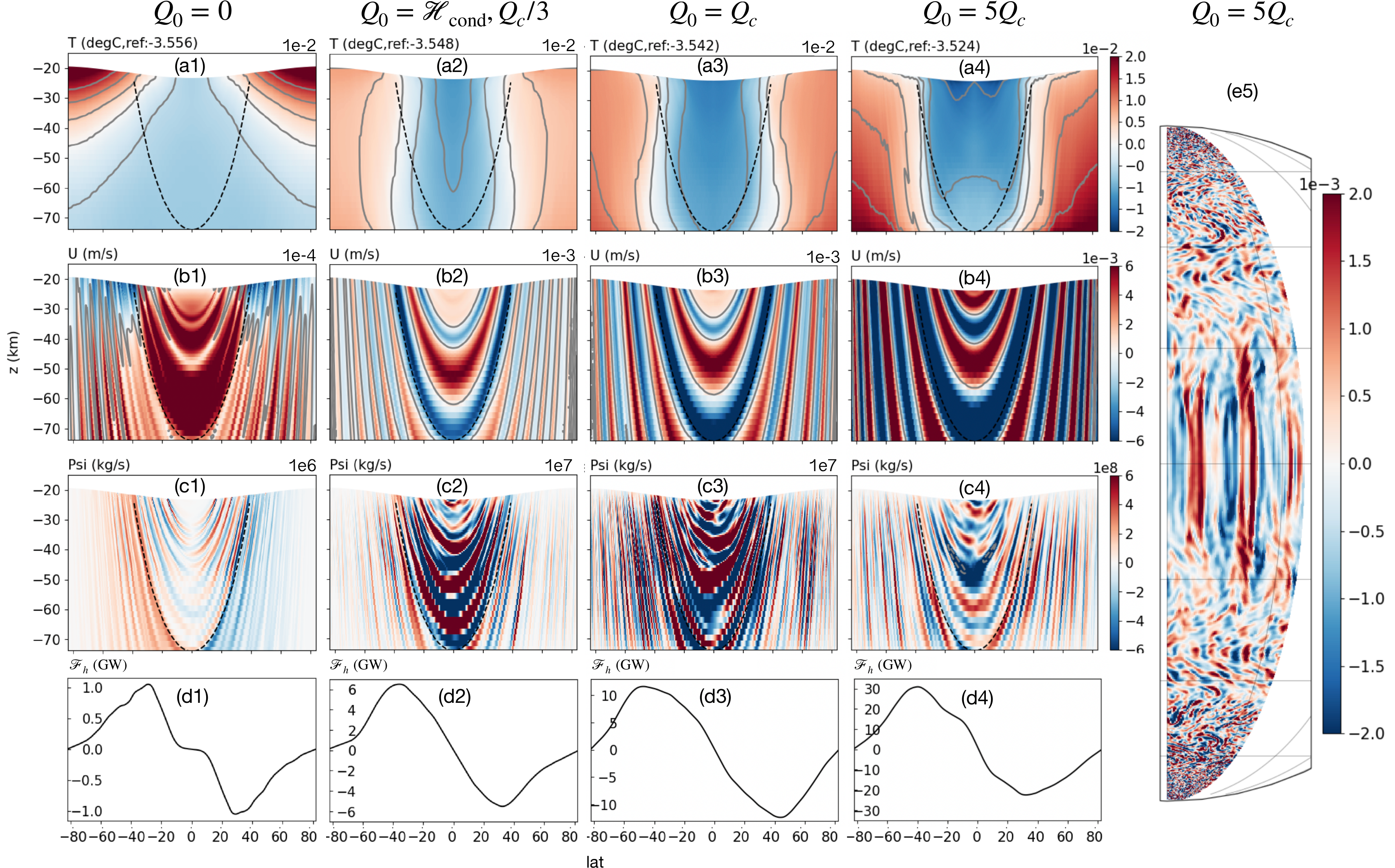}
    \caption{\small{Solutions for the various heating scenarios with $a=250$~km. The top three rows show zonally-averaged temperature $T$, zonal flow speed $U$, meridional streamfunction $\Psi(\phi,z) = \int_{-H_{\mathrm{tot}}}^z  \rho(\phi,z') V(\phi,z')\times (2\pi(a+z')\cos\phi)~ dz'$, where $\phi$ denotes latitude, $z$ denotes altitude, $-H_{\mathrm{tot}}$ is the altitude of the seafloor and $V$ and $\rho$ are meridional speed and density. $\Psi>0$ indicates clockwise circulation. The last row (panel d1-4) shows the vertically-integrated meridional ocean heat transport $\mathcal{F}(\phi) = \int_{-H_{\mathrm{tot}}}^{-H}  \rho(\phi,z') V(\phi,z')T(\phi,z')\times (2\pi(a+z')\cos\phi)~ dz'$. $\mathcal{F}>0$ if heat is transport northward. From the left to the right column show solutions with $Q_0=0$, $Q_0=\mathcal{H}_{\mathrm{cond}}f_{\mathrm{deep}}\approx Q_c/3$, $Q_0=Q_c$ and $Q_0=5Q_c$, respectively, where $Q_c=0.21$~W/m$^2$. }}
    \label{fig:solution-a250}
  \end{figure*}

\begin{figure*}
    \centering \includegraphics[page=7,width=\textwidth]{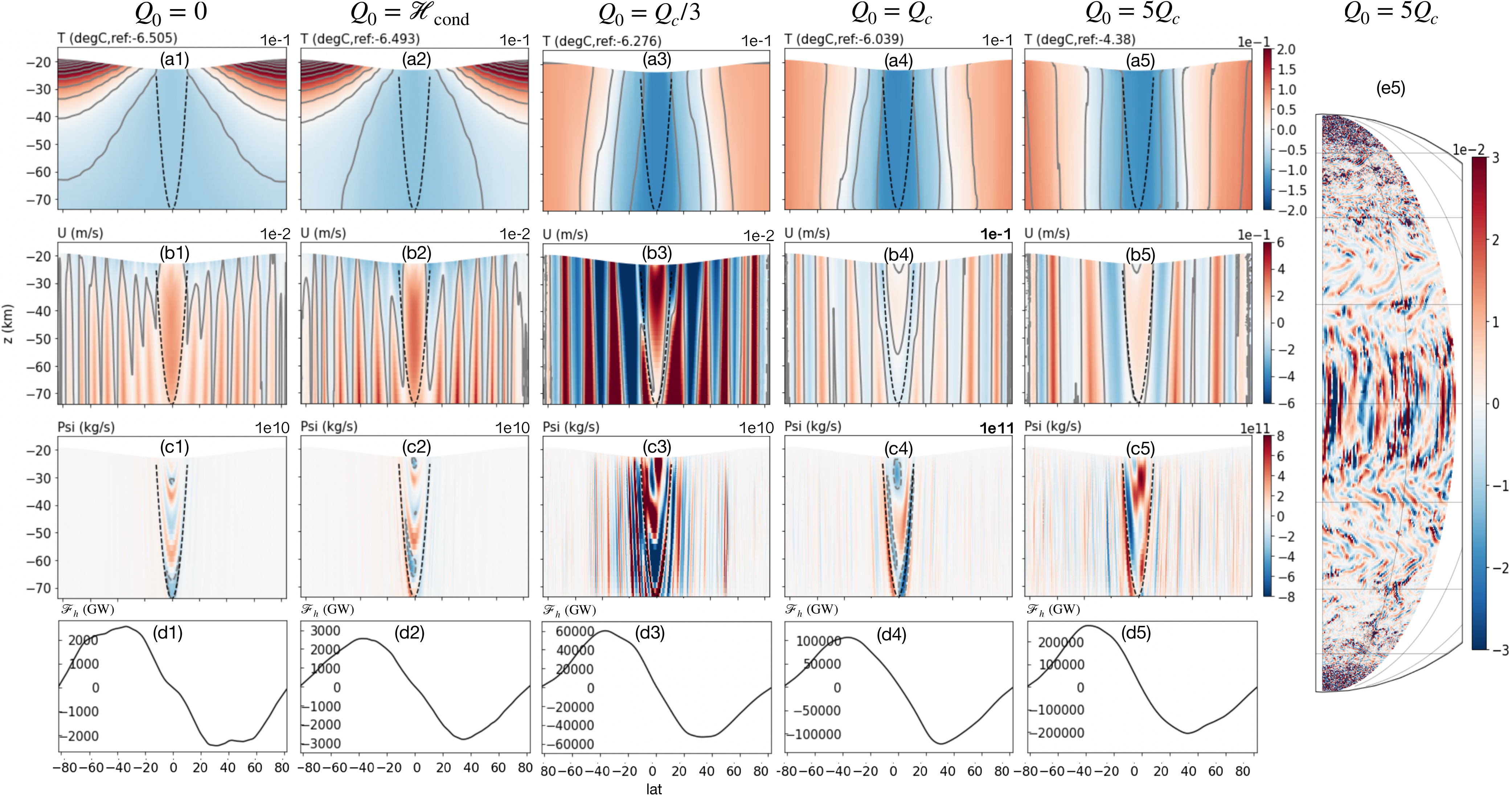}
    \caption{\small{Same as Fig.~\ref{fig:solution-a250} except satellite radius $a=2500$~km. Here $Q_c=17.7$~W/m$^2$. }}
    \label{fig:solution-a2500}
  \end{figure*}

Fig.~\ref{fig:solution-a250} and Fig.~\ref{fig:solution-a2500} show the same info as Fig.~2 in the main text for the experiments with $a=1000,\ 2500$~km radius.
  
  \section{Sensitivity tests.}
  \label{sec:sensitivity}
\begin{figure*}
    \centering \includegraphics[page=8,width=0.5\textwidth]{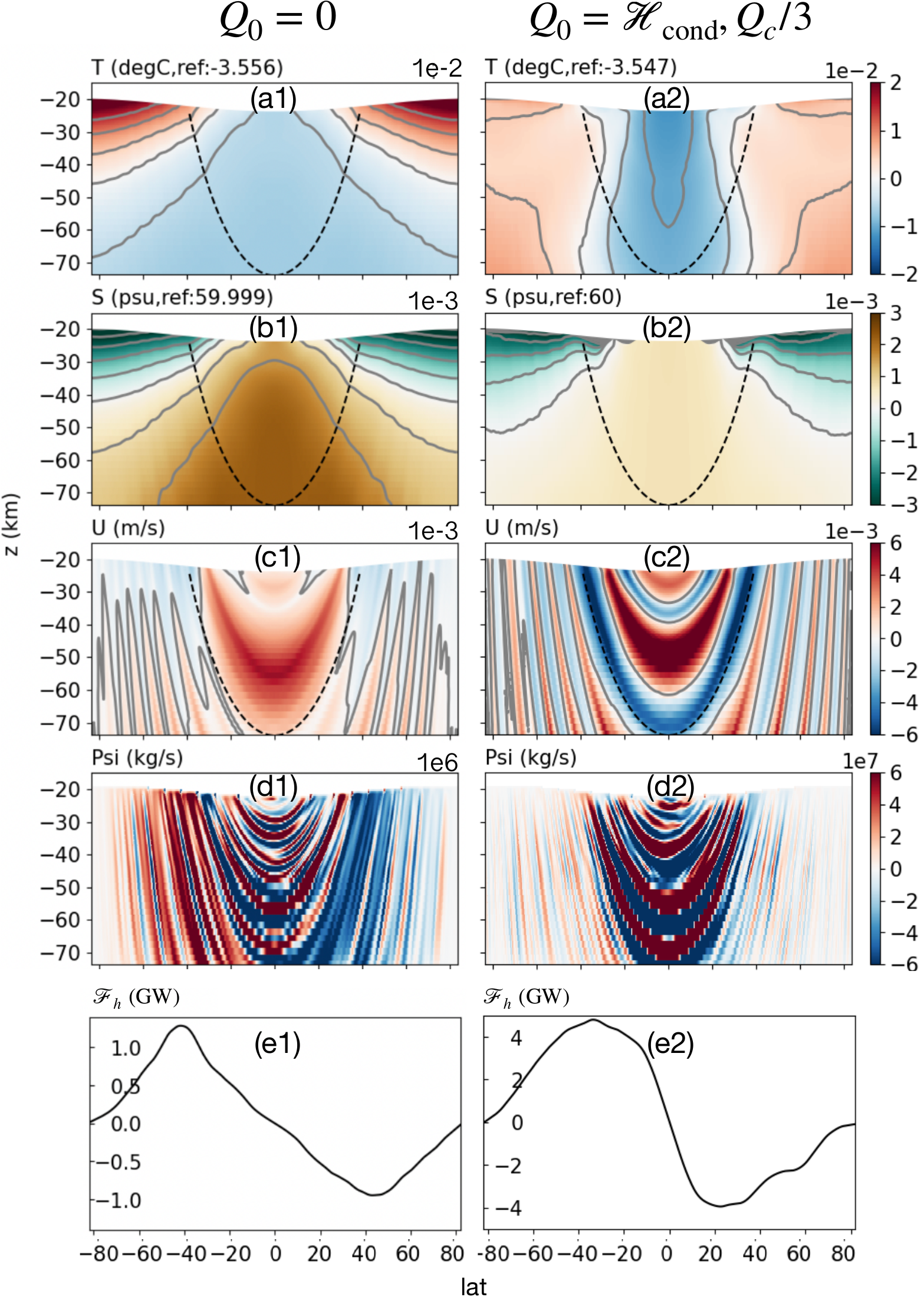}
    \caption{\small{Same as Fig.~\ref{fig:solution-a250} text except nonlinear equation of state ``MDJWF'', which accounts for both temperature and salinity induced density change, is adopted. $a=250$~km. Only the core-heating and shell-heating scenarios are repeated in the sensitivity test here. }}
    \label{fig:solution-a250-mdjwf}
  \end{figure*}

  Fig.~\ref{fig:solution-a250-mdjwf} shows the solutions of the sensitivity tests for nonlinear equation of state and the salinity-driven circulation for $a=250$~km. Comparison should be made against the first two columns of Fig.~2 in the main text.

\begin{figure*}
    \centering \includegraphics[page=9,width=0.7\textwidth]{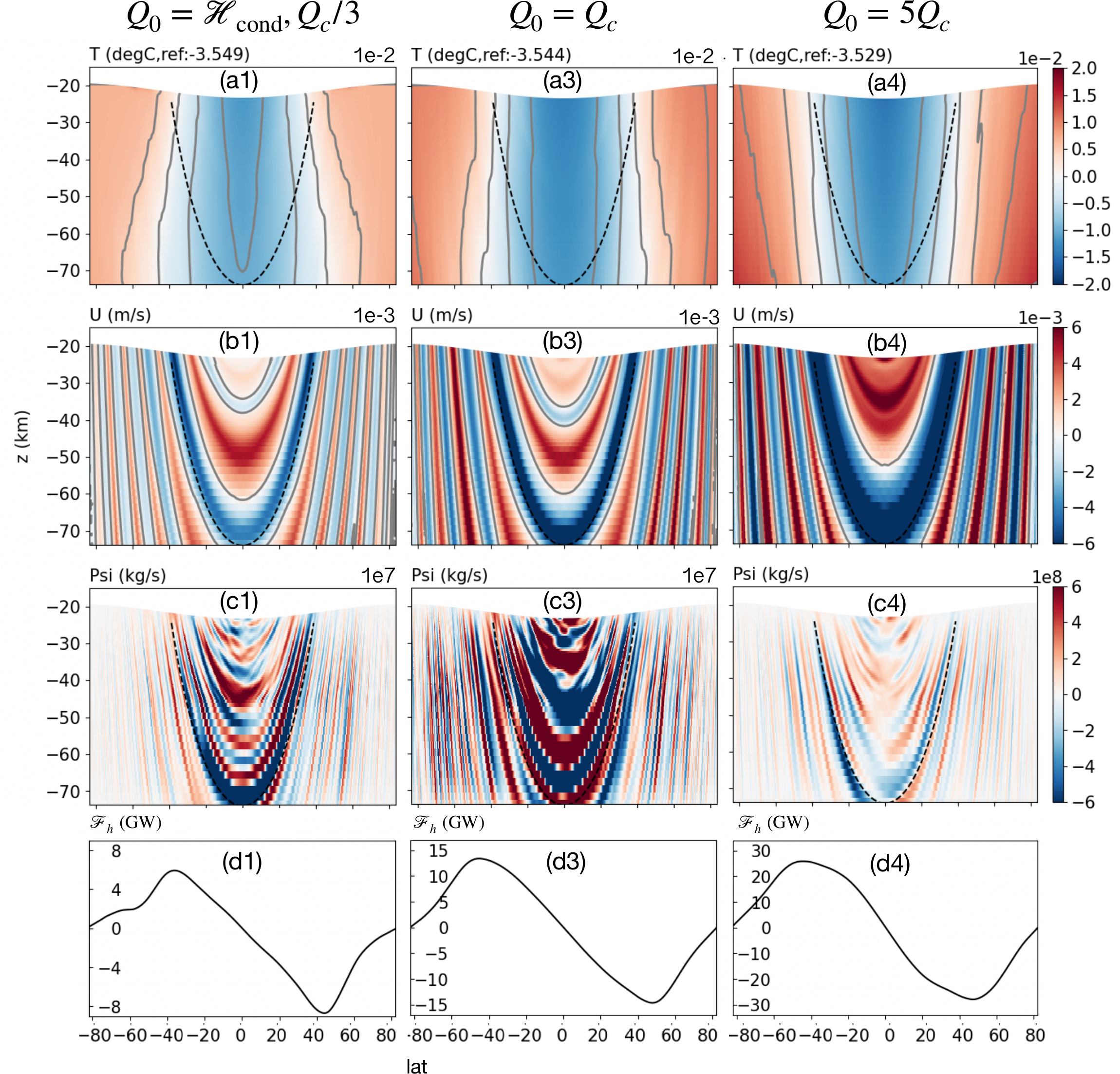}
    \caption{\small{Same as Fig.~\ref{fig:solution-a250} except convective parameterization is on. From left to right the bottom heating $Q_0=Q_c/3,\ Q_c$ and $5Q_c$, respectively. }}
    \label{fig:solution-a250-kconv}
  \end{figure*}

  Fig.~\ref{fig:solution-a250-kconv} shows the solutions of the sensitivity tests for convective parameterization, which works by boosting the vertical diffusion in regions where stratification becomes unstable to $2$ times the value given by Eq.6 in the main text. Only the three strongly forced experiments are repeated here because they are likely to be more affected by convective parameterization. As can be seen by comparing Fig.~\ref{fig:solution-a250-kconv} with Fig.3 in the main text, convective parameterization weakens the vertical temperature gradient induced by bottom heating, but this does not substantially affect the meridional ocean heat transport. 
  
\begin{figure*}
    \centering \includegraphics[page=12,width=0.7\textwidth]{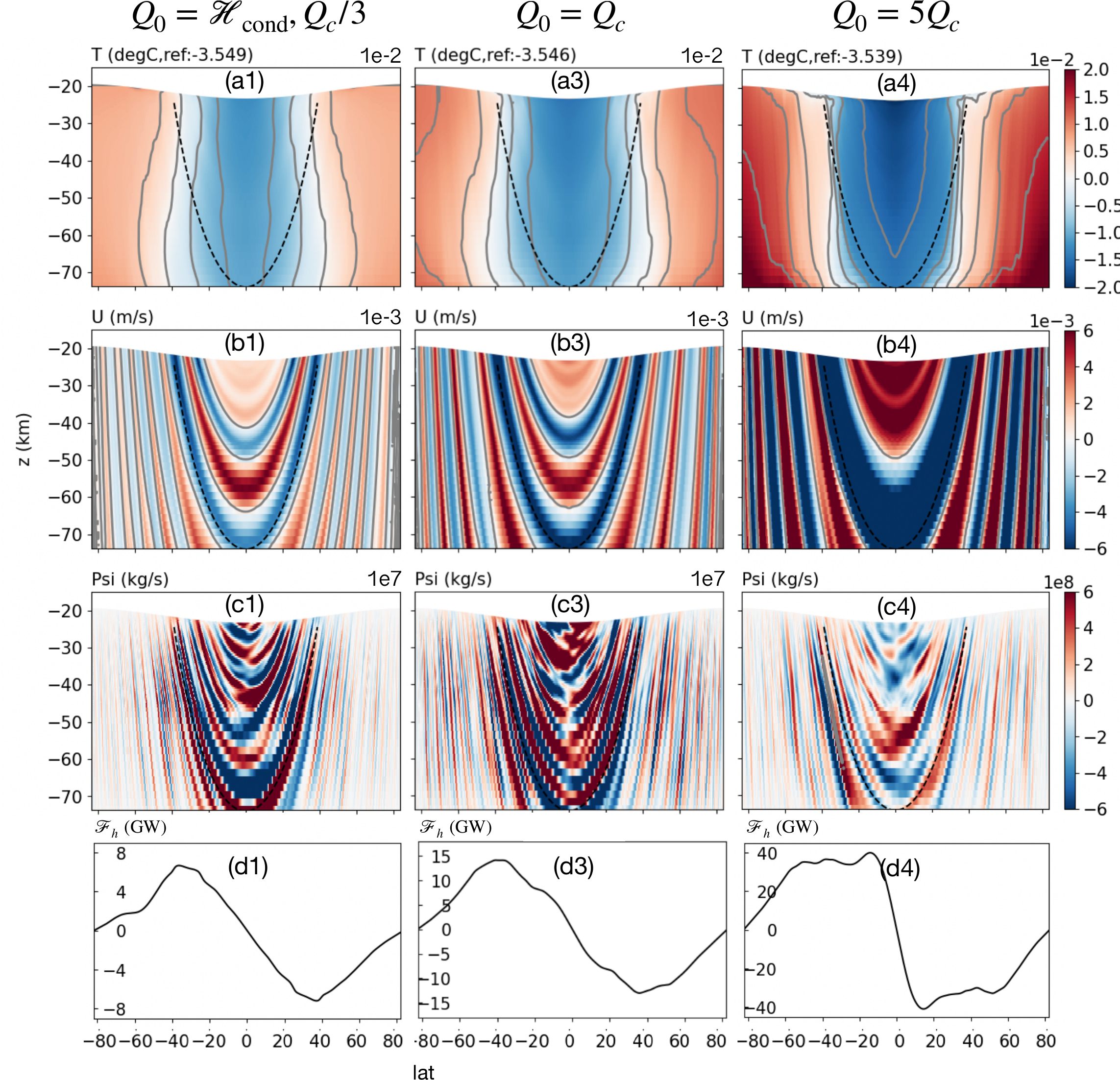}
    \caption{\small{Same as Fig.~\ref{fig:solution-a250} except the boundary heat/salt exchange coefficient $\gamma_T, \gamma_S$ is enhanced by a factor of 50, which may be relevant due to the strong ocean currents in these strongly forced experiments. From left to right the bottom heating $Q_0=Q_c/3,\ Q_c$ and $5Q_c$, respectively. }}
    \label{fig:solution-a250-gmTS50}
  \end{figure*}

\begin{figure*}
    \centering \includegraphics[page=11,width=0.7\textwidth]{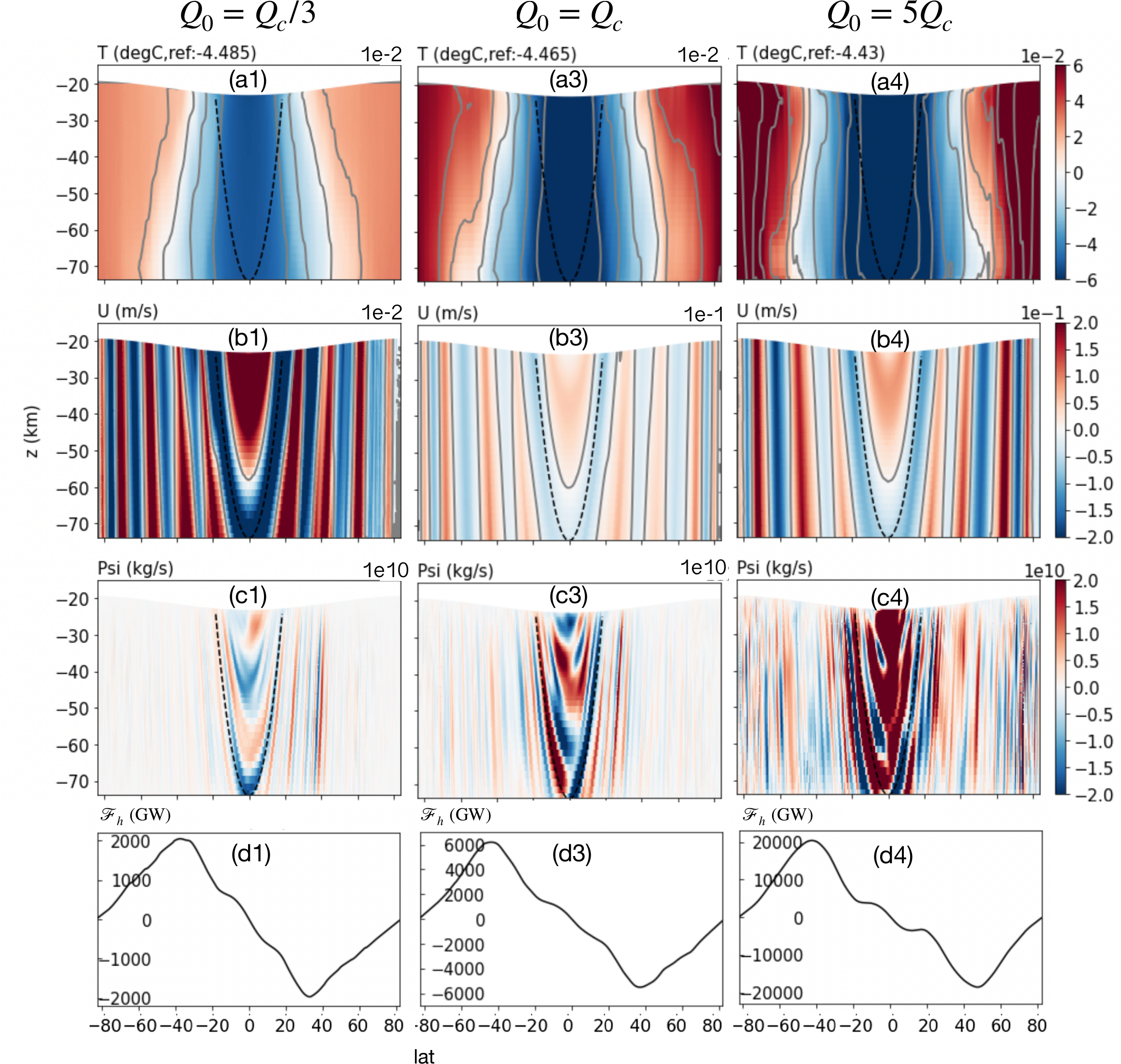}
    \caption{\small{Same as Fig.3 in the main text except the boundary heat/salt exchange coefficient $\gamma_T, \gamma_S$ is enhanced by a factor of 50, which may be relevant due to the strong ocean currents in these strongly forced experiments. From left to right the bottom heating $Q_0=Q_c/3,\ Q_c$ and $5Q_c$, respectively. }}
    \label{fig:solution-a1000-gmTS50}
  \end{figure*}

\begin{figure*}
    \centering \includegraphics[page=10,width=0.7\textwidth]{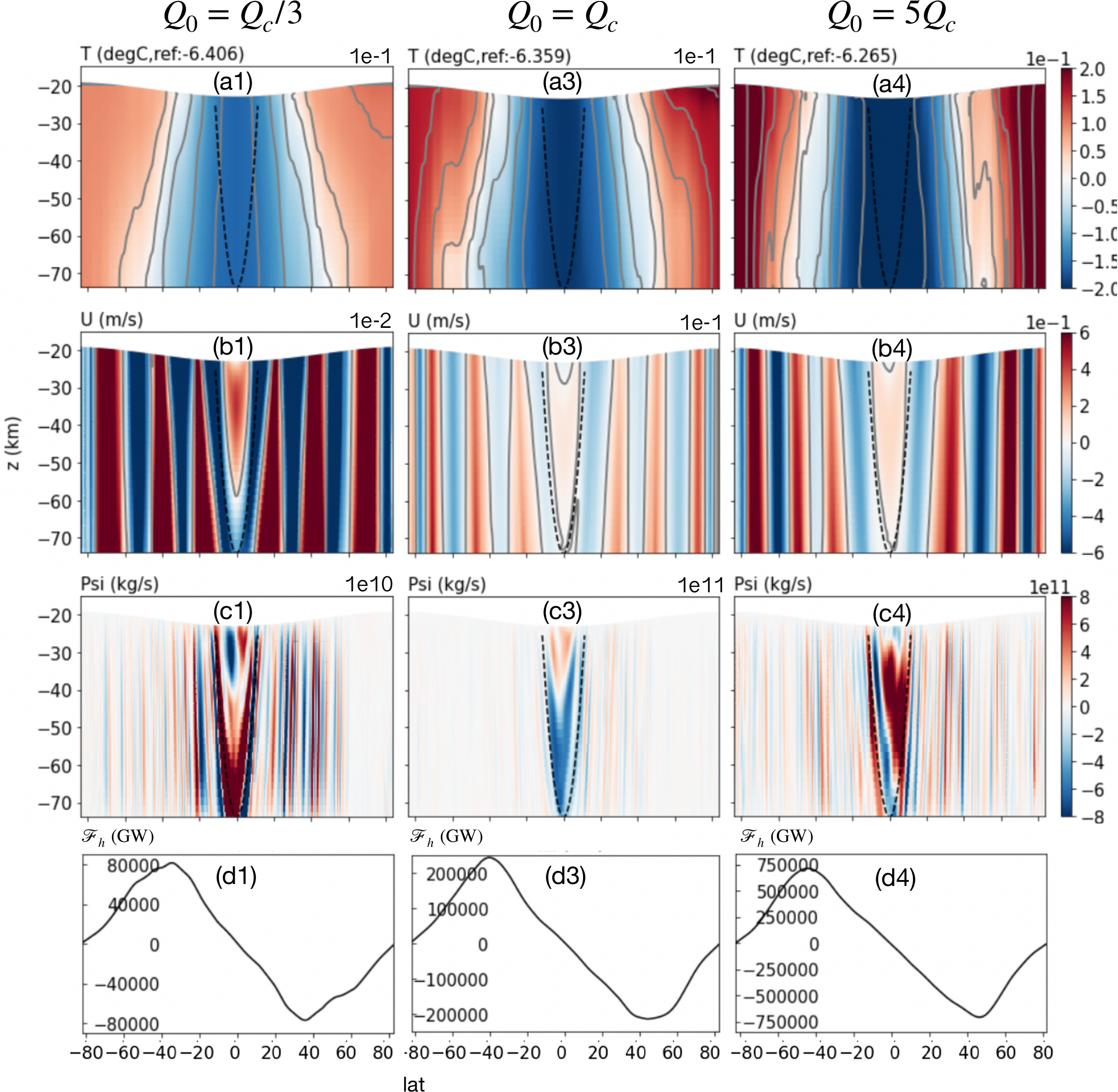}
    \caption{\small{Same as Fig.~\ref{fig:solution-a2500} except the boundary heat/salt exchange coefficient $\gamma_T, \gamma_S$ is enhanced by a factor of 50, which may be relevant due to the strong ocean currents in these strongly forced experiments. From left to right the bottom heating $Q_0=Q_c/3,\ Q_c$ and $5Q_c$, respectively. }}
    \label{fig:solution-a2500-gmTS50}
  \end{figure*}

  Fig.~\ref{fig:solution-a2500-gmTS50} shows the solutions of the sensitivity tests for boundary heat/salt exchange coefficient $\gamma_T, \gamma_S$. By default, $\gamma_T, \gamma_S$ are set to a very small value, $10^{-5}$~m/s, which is appropriate when the ocean currents are weak. However, the strongly forced cases may have rather strong ocean currents, and underestimating $\gamma_T, \gamma_S$ will block the heat from being delivered to the equatorial ice shell. In order to deliver $2Q_0$ of heat flux to the ice shell, the ocean temperature beneath the thick ice needs to be higher than the freezing point by $2Q_0/(\rho C_p \gamma_T)$. This reduces the meridional temperature gradient under the ice $\Delta_hT$, as can be seen in Fig.~\ref{fig:solution-a250}, Fig.~\ref{fig:solution-a2500} and Fig.3 in the main text. When $\gamma_T, \gamma_S$ is enhanced by a factor of $50$, the reduction of $\Delta_hT$ significantly reduces (see Fig.~\ref{fig:solution-a250-gmTS50}-\ref{fig:solution-a2500-gmTS50}), which then facilitates equatorward heat deflection (Fig.4 in the main text).

    \begin{figure*}
    \centering \includegraphics[page=13,width=0.7\textwidth]{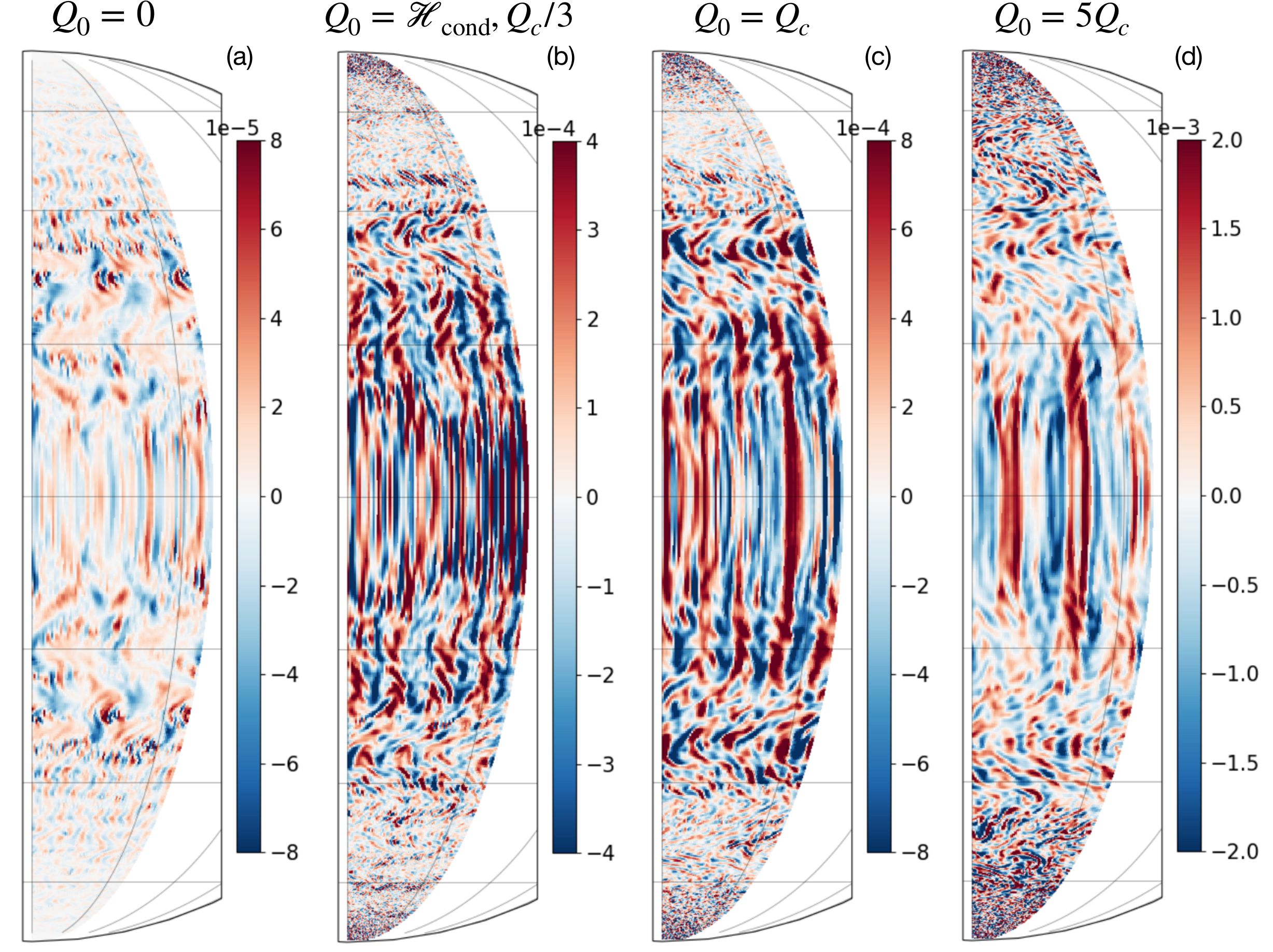}
    \caption{\small{Mid-level vertical flow field for a=250~km scenarios.}}
    \label{fig:Whori-a250}
  \end{figure*}

      \begin{figure*}
    \centering \includegraphics[page=14,width=0.7\textwidth]{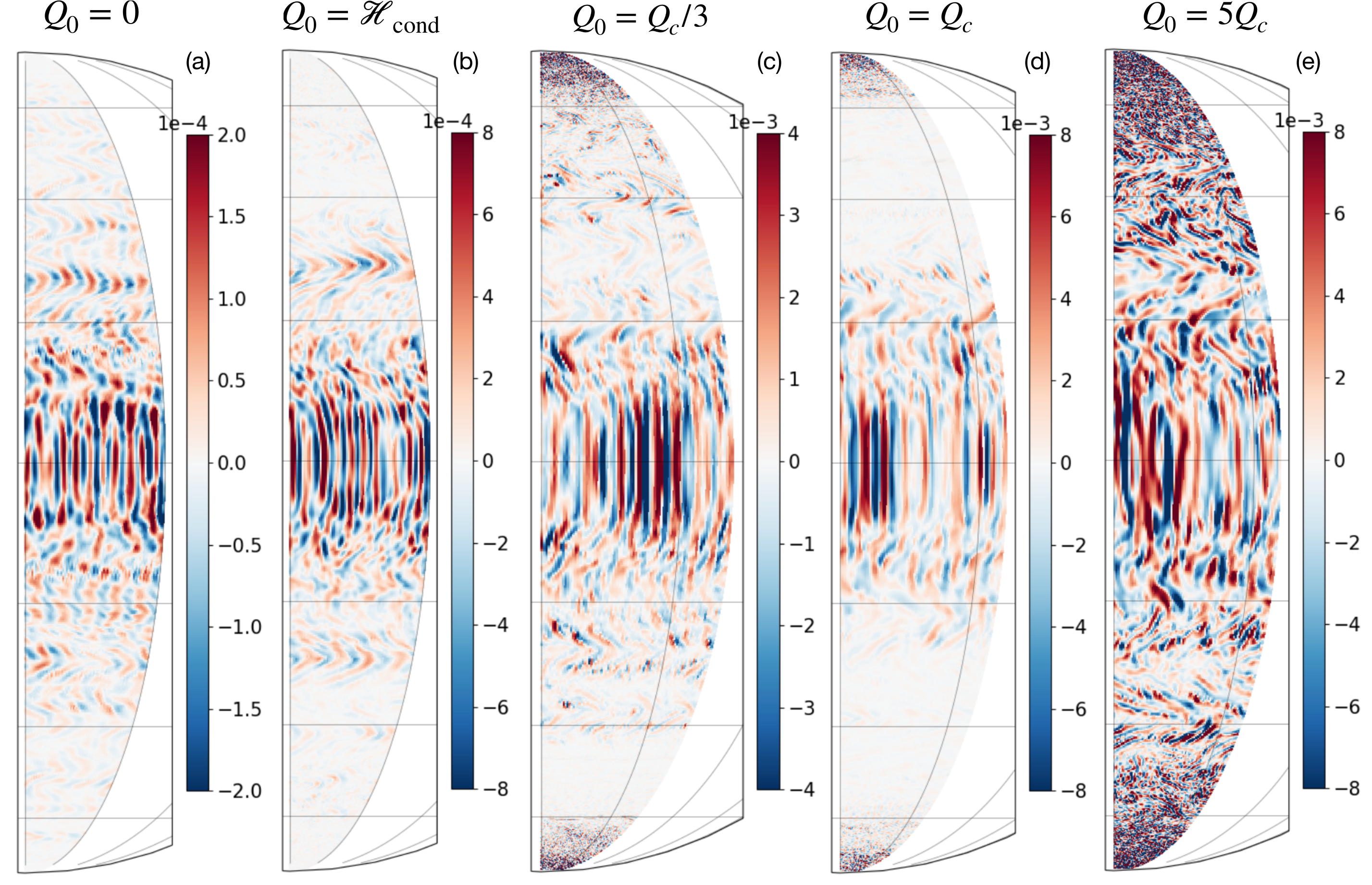}
    \caption{\small{Mid-level vertical flow field for a=1000~km scenarios.}}
    \label{fig:Whori-a1000}
  \end{figure*}

        \begin{figure*}
    \centering \includegraphics[page=15,width=0.7\textwidth]{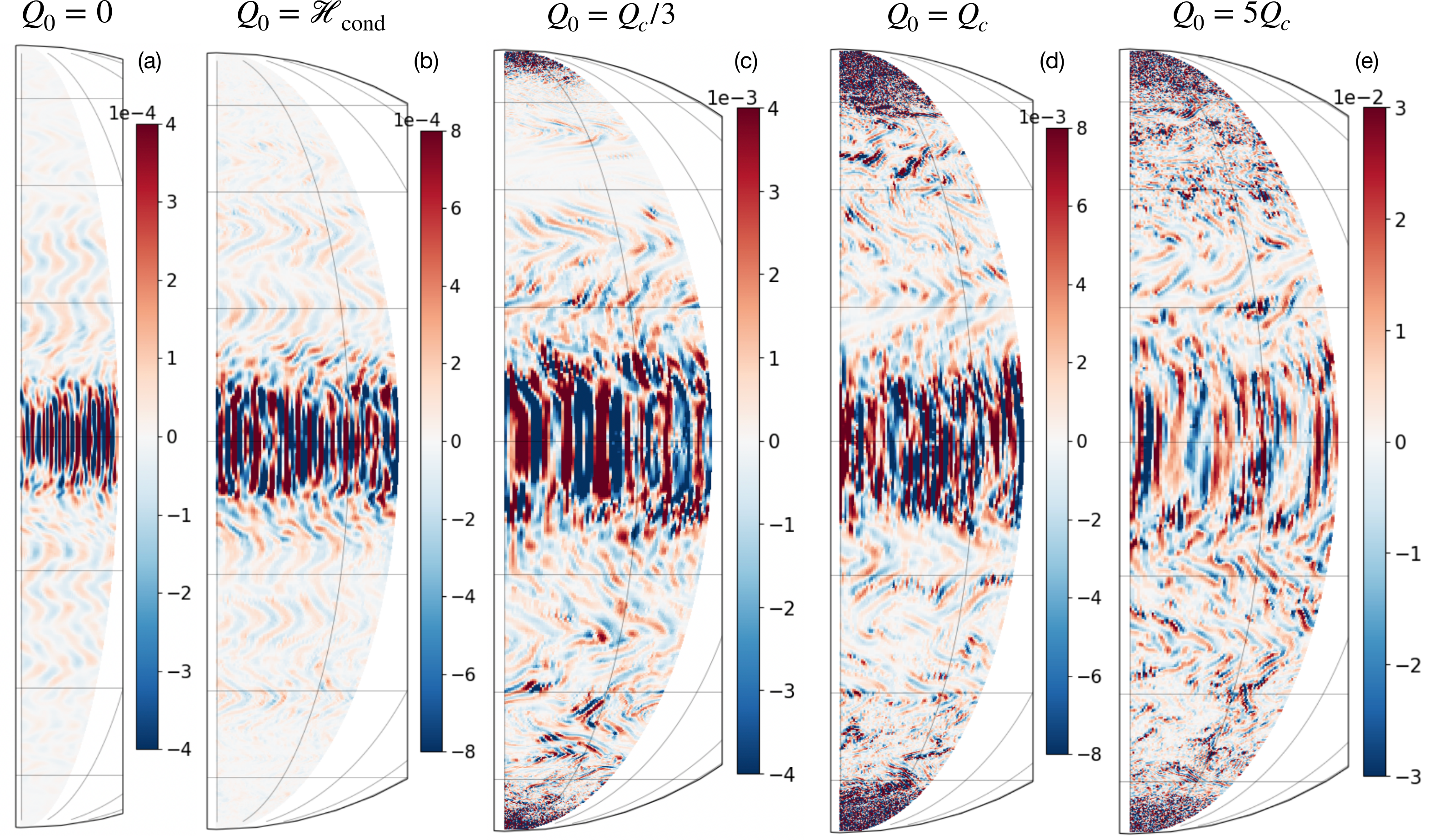}
    \caption{\small{Mid-level vertical flow field for a=2500~km scenarios.}}
    \label{fig:Whori-a2500}
  \end{figure*}


\bsp	
\label{lastpage}
\end{document}